\documentclass[12pt]{iopart}

\usepackage{graphicx}
\usepackage{bm}        

\begin{document}
\newcommand{\braket}[3]{\bra{#1}\;#2\;\ket{#3}}
\newcommand{\projop}[2]{ \ket{#1}\bra{#2}}
\newcommand{\ket}[1]{ |\;#1\;\rangle}
\newcommand{\bra}[1]{ \langle\;#1\;|}
\newcommand{\iprod}[2]{\bra{#1}\ket{#2}}
\newcommand{\logt}[1]{\log_2\left(#1\right)}
\def\cI{\mathcal{I}}
\def\cC{\tilde{C}}
\newcommand{\cx}[1]{\tilde{#1}}

\def\be{\begin{equation}}
\def\ee{\end{equation}}
 \title[signal of a local decohering processs in spin chains]{Remotely detecting the signal of a local decohering process in spin chains} 
\author{Saikat Sur$^1$ and V. Subrahmanyam$^2$}
 \address{ Department of Physics, Indian Institute Of Technology,  Kanpur-208016, India}
 
 \ead{saikatsu@iitk.ac.in$^1$ and vmani@iitk.ac.in$^2$}
\date{\today}
\begin{abstract}
We study the dynamics of a one dimensional quantum spin chain evolving from
unentangled or entangled initial state. At a given instant of time a quantum
dynamical process (ex. measurement) is performed on a single spin at one end of
the chain,  decohering the system. Through the further unitary evolution, a signal propagates in the spin chain,  which can be detected from a measurement on a different spin at later times. 
From the dynamical unitary evolution of the decohered state from the epoch time,
it is possible to detect the occurrence of the dynamical process. The propagation of the signal for the dynamical process, and the speed of the signal are
investigated for various spin models, viz. using the Ising, Heisenberg, and the transverse-XY dynamics.
\vskip 0.5cm
\hfill {PACS: 03.65.Ud 03.67.Bg 03.67.Hk 75.10.Pq}
\end{abstract}
\maketitle

 \maketitle
\section{ Introduction}

Quantum information and communication aspects of quantum spin chains, which can be viewed as multipartite systems of qubits, have been investigated over the last few years, as a spin chain is a possible channel for quantum
state transfer\cite{ref1,ref2,ref3}.
Quantum spin chains have been studied extensively as prototype condensed matter systems that exhibit quantum critical behaviour, novel spin states with a variety of spin ordering\cite{ref5,ref6,ref7}. These systems are generally studied from the quantum dynamics view point, i.e. the evolution of an initial quantum many-body state through the time-dependent Schroedinger equation\cite{ref9}, and from a statistical mechanics view point, i.e. the various thermodynamic phases and transitions\cite{ref5,ref9}. 

The dynamics of spin chains have
been investigated for magnon bound states and scattering\cite{refa1,refa2}, spin current dynamics\cite{refc2}, relativistic density wave dynamics\cite{refc3}. The evolution of quantum correlations  after a quantum quench\cite{refb1}, the light-cone in entanglement spreading\cite{refb2} have  been studied using the dynamics of  model Hamiltonians. 
All these studies involve the unitary evolution using the Schroedinger equation of initial chosen states or after a quench, and subsequent redistribution of quantum correlations.
Now,  from the quantum information theory view point, a many-qubit state can undergo various multi-qubit gate operations, both global and local unitaries, and will undergo redistribution of 
quantum correlations, e.g. entanglement\cite{ref10,ref11, ref13,ref14,ref15} among the various
subsystems, but as a whole the multi-qubit system is treated as closed system. However, a multi-qubit system that is used in any quantum communication protocol will become an open
system, as various subparts can undergo non-unitary operations, i.e. a general quantum dynamical process (QDP) or a quantum channel action\cite{ref16,ref17}. This is a common source of decoherence\cite{ref18},
which is a stumbling block for a faithful communication of quantum states. 

In this paper, we consider  a quantum spin chain that undergoes decoherence, from a QDP that occurs at a given spin.  We examine the scenario of a QDP signal propagating through
the spin chain, and  detecting it from another spin a distance away, whether the QDP occurred or not. As the quantum dynamics is due to a Hamiltonian evolution from local interactions, the efficiency of detection falls as we move away from the site of QDP occurrence. 
Moreover, it is expected that further sites will take longer time to detect the occurrence, as the speed of communication of the fact that QDP occurred will
be determined by various details, and similarly for the efficiency of detecting the QDP.  
Intuitively we can say that, if the spins interact with their nearest neighbours, if any quantum dynamical process occurred at the boundary spin, its signal  can only propagate with a finite speed through the chain.  So, the observer will have to wait for some time before detecting the signal. But in the case of a long range interaction we expect a larger speed. The speed of propagation of correlations in spin chains for the case of unitary dynamics, and its dependence on the model parameters has been studied\cite{refa1,refb1}. Similarly in this case of
non unitary dynamics,  it is expected that the speed will depend on several factors, the type of the interaction, strength of the interaction, the spatial range of interaction strength, the initial state of the system, the external magnetic field strength. However, the speed of detection may not have the same dependence on the model parameters, as we shall see below.

This paper is arranged as follows. In the next section we describe the general approach and a numerical algorithm for the dynamical evolution of the state, for any type of Hamiltonian that
governs the dynamics. We setup a  detector function for a simple QDP, namely a known projective measurement process on a given spin. In  Section III, we investigate the
simplest of the evolution, that is Ising dynamics, and investigate the signal propagation and detection of QDP both analytically and numerically. We consider both entangled and non-entangled initial states. In Section IV and V, we analytically calculate the detector function for specific state for Heisenberg and transverse-XY models respectively, along with numerical results. In each of these
cases, the dynamics can be investigated analytically for a spin chain, as these models admit of exact solutions for all many-spin eigenstates and eigenvalues, through Bethe Ansatz \cite{ref7}, and
Jordon-Wigner transformation respectively\cite{ref19,ref20}.In these cases, as we shall see, there is a finite speed of signal propagation giving a definite possibility of detecting the QDP signal. Finally we summarise our results in Section VI.  

\section{{  Signal from a QDP and its detection in a multi-party system }}

A multi-partite system can exhibit a variety of correlations among its various many parts. In a general multi-qubit pure state of the system, there can be pairwise quantum correlations\cite{ref21,ref22} between two parts
A and B. Let us consider a situation where there is a QDP that occurs on part A, which leads to decoherence.  Now, the question is whether this fact that the QDP occurred on part A can be detected from spatially separated part B. No communication or No signalling theorem \cite{ref16,ref24,ref25, ref26, ref27} says that it is impossible, notwithstanding the preexisting quantum correlations, unless there is  a further evolution of the state from the epoch of QDP occurrence to the epoch of detecting it from B.

\begin{figure}
\center{
 \includegraphics[width=0.75\textwidth]{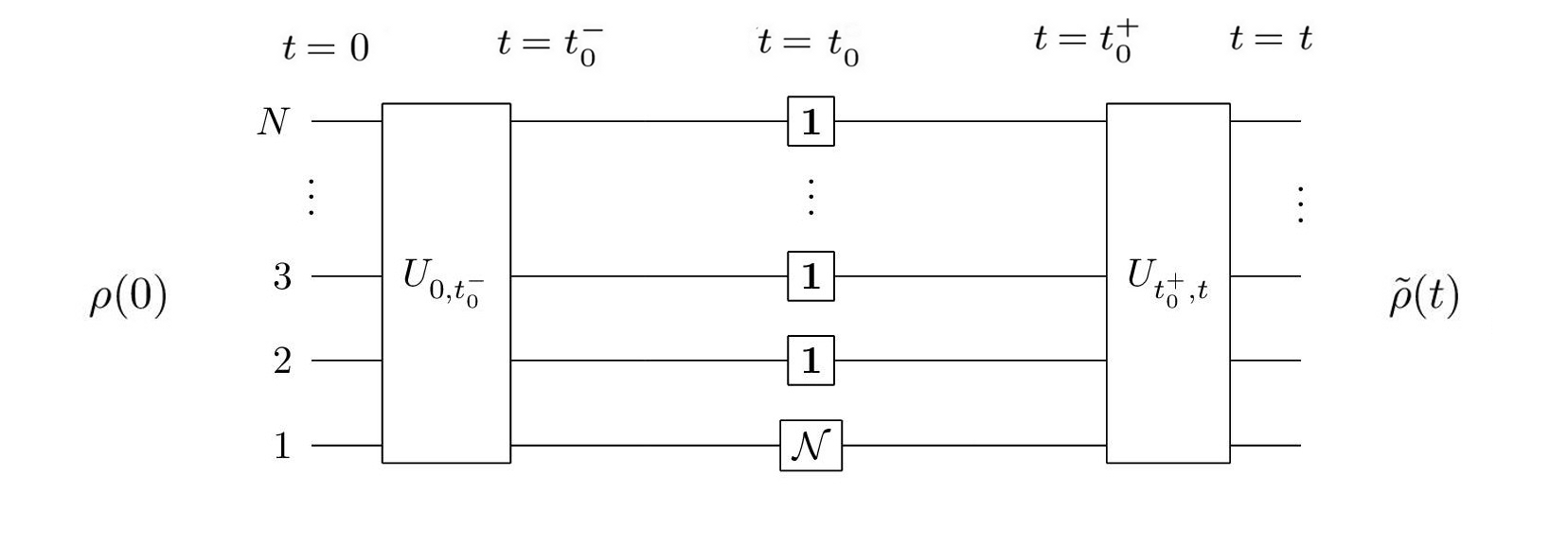} }
 
\caption{ \small 
Fig. 1 A quantum circuit describing the sequence of operations,  a global unitary transformation from the Hamiltonian dynamics, the instantaneous local channel action on the first qubit, followed by a  subsequent unitary operation. The Hamiltonian evolution of the initial state is  interrupted with a local quantum channel action on the first qubit, which can be detected from a different qubit}

\end{figure}

To see this for a multi-qubit system, let us consider a linear chain of $N$ spins, with an initial state $\rho$. The details of the structure of the correlations are not important for arguing the no signalling theorem. Let the first qubit undergo a QDP,  a general instantaneous evolution that includes decohering process. This amounts to a quantum channel action of the first qubit.
 The many-qubit state is instantaneously transformed into a state $\tilde \rho$ through a quantum channel action
${\cal N}_1$ on the first qubit. In this evolution, the operation transforms the input state  into an output state through the quantum channel action, ${{\cal N}_1\times {\cal I}_2\times..{\cal I}_N}$. That is only the first qubit undergoes the QDP, and the rest of the qubits are operated by the identity operator.
This QDP  results in an output density matrix, viz. 
\begin{equation}
\rho~~ \stackrel {{\cal N}_1\times {\cal I}_2\times..{\cal I}_N}{{ \longrightarrow}}~~\tilde \rho =\sum_i P_i\rho P_i^\dag,
\end{equation}
where we have used the Kraus operators $\{P_i\}$ for the quantum channel\cite{ref16,ref17}, with the constraint that $\sum P_i^\dag P_i=1$.
Now, we would like to detect this QDP from a different qubit from a measurement on $n'$th qubit, which depends only on the reduced density matrix for the desired qubit.
However, it is not possible to detect from any other qubit whether or not QDP occurred on the first qubit, due to the No Communication Theorem. This is can easily be seen by comparing
the two reduced density matrices for the desired qubit, without and with the QDP, given by
\begin{equation}
\rho_n= Tr^\prime \rho,~ \tilde \rho_n = Tr^\prime {\tilde \rho},
\end{equation}
where the prime indicates a partial trace over all qubits other than $n'$th qubit. Any measurement done on the desired qubit will involve an operator $A_n$ that depends on the spin operators for the $n'$th qubit. We can see that the two reduced density matrices will give identical results for the expectation value,
\begin{equation}
\langle A_n\rangle = Tr A_n \tilde \rho = \sum_i Tr A_n P_i \rho P_i^\dag = Tr A_n\rho.
\end{equation}
In the last step, we have used the fact that $P_i$ commutes with $A_n$ as $P_i$ depends only on spin operators of the first qubit, and the completeness relation
$\sum P_i^\dag P_i=1$. Thus, it is not possible to distinguish between the two reduced density matrices, implying that it is not possible to detect from other qubits whether the QDP occurred or not at the first qubit. This is because, we have not considered the evolution of the state after the QDP occurs and before the detection. Below, we will include the further evolution of the state, which will enable the detection of QDP. We will also define a detector function to quantify the efficiency.

Let us consider the scenario of an initial state $\rho(t=0)=|\psi(0)\rangle\langle \psi(0)|$, a pure state, evolving through a Hamiltonian evolution, with given Hamiltonian $H$ that includes pairwise spin interactions.  
We can write the state using the basis states
$|0\rangle$ and $|1\rangle$, the eigenstates of $\sigma^z$ operators of the individual qubits, we have
\begin{equation}
|\psi(0)\rangle = \sum \psi_{s_1,s_2..s_N}|s_1,s_2..s_N\rangle,
\end{equation}
where the sum is over qubit states, taking $s_i=0,1$ for the $i'$th qubit. 
The different basis states are classified into odd and even magnon states. Even (odd) magnon states have even (odd) number of qubit flips from the ferromagnetic state $|00..0\rangle$.
For simplicity, in the beginning we consider initial states with even-only (or odd-only) magnon basis states with nonzero wave function amplitudes, i.e. only even (or only odd) number of qubits with $s_i=0$. Later we will
discuss the case of mixing the even and odd sector states.
The Hamiltonians for the unitary dynamics that we consider in the later sections, both the
Heisenberg and transverse-XY models conserve the evenness (or oddness) of the number of qubits in the state $|0\rangle$.  The Hamiltonian unitary evolution of the initial states involves
magnon excitations. We employ the periodic boundary conditions in most of the cases, for finding the excitation spectrum and the eigenfunctions analytically. For numerical calculations
we employ open boundary conditions.
Now, the initial state undergoes transformation through a sequences of operations as shown in Fig. 1. First, there is a continuous unitary evolution from $t=0$ to $t=t_0^-$. At $t=t_0$, the
state now is given by
\begin{equation}
\rho(t_0^-)= U_{0,t_0} \rho U_{0,t_0}^\dag.
\end{equation} In the second step, the system undergoes a QDP, 
a quantum channel action on the first qubit instantaneously. Let us consider a simple projective measurement done on the first qubit, in the eigen basis of the operator $\vec \sigma_1.\hat n$. The Kraus operators for this QDP are given
by,  $P_0=(1+\vec \sigma_1.\hat n)/2$, and $P_1=(1-\vec \sigma_1.\hat n)/2$, corresponding to a  measurement process (that measures $\vec \sigma_1.\hat n$) on the first qubit. Now, the resultant state, immediately after the
QDP occurrence, is written as,

  \begin{equation}
\tilde{{\rho}}(t_{0}^{+}) = P_{0}{\rho}(t_{0}^{-})P_{0}^{\dagger} + P_{1}{\rho}(t_{0}^{-})P_{1}^{\dagger}.
  \end{equation}
 In the third step, the state is further evolved to a time $t>t_{0}$. Now, the final state $\tilde \rho(t)$ is then given by,
 $\tilde{{\rho}}(t)= U_{t,t_{0}} \tilde \rho(t_0^+) U_{t,t_0}^\dag,$
 is given by,
 \begin{eqnarray}
 \tilde{{\rho}}(t)&=& \frac{1}{2}{\large [} U_{t,0}{\rho}(0)U_{t,0}^{\dagger} +  U_{t,t_{0}}{\vec \sigma_1}.\hat nU_{t_{0},0}{\rho}(0)U_{t_{0},0}^{\dagger}{\vec \sigma_1}.\hat n U_{t,t_{0}}^{\dagger} {\large ]} ,\nonumber \\
 &=&{1\over 2} \lbrace |\psi(t)\rangle\langle \psi(t)|  + |\tilde \psi(t)\rangle\langle \tilde\psi(t)| \rbrace.
  \end{eqnarray} 
The state is written in terms of two pure states, $|\psi(t)\rangle\equiv e^{-iHt}|\psi(0)\rangle$ that carries the time evolution without reference to the QDP occurring at $t_0$, and
$|\tilde \psi(t)\rangle \equiv e^{-iH(t-t_0)} \vec \sigma_1.\hat n |\psi(t_0)\rangle$ that caries the effect of the QDP occurrence. 
The first term in the above is just $\rho(t)$, the state at time $t$ with no QDP occurring at the first site at time $t=t_0$. This would have been the evolved state, if the decohering QDP did
not occur, with a smooth unitary evolution. The second term has the information regarding
the decohering process, through the time evolution from $t=t_0$ onwards. It should be emphasised here that the two states, $\tilde \rho(t)$ and $\rho(t)$, with and without the local
instantaneous QDP intervening the unitary dynamics, will be differ only slightly. This is due to the fact that only one among $N$ spins is affected upon by the QDP at the epoch time
$t_0$, and the state is evolved further with the unitary dynamics. Thus, these two states only differ to this extent, and both states are further evolved through the same Hamiltonian dynamics. The spread of correlations would also differ to that much extent only. The model parameter dependence of the evolution of correlations are expected to be similar for
both these states. Thus, we expect the efficiency of detecting the QDP  from a different qubit will have much weaker dependence on the model parameter. We shall see
in the following sections, using different model Hamiltonian dynamics, the possibility of detecting the QDP signal.

Now, it is possible to detect the QDP from a measurement on $n$'th qubit, as the reduced density matrix  $\tilde \rho_n = Tr^\prime \tilde \rho(t)$ can be differentiated from the reduced density $\rho_n = Tr^\prime  \rho(t)$.  The reduced density matrix for the $n$'th site, after tracing out other spins, is given in the diagonal basis of $\sigma_n^z$ by
\begin{equation}
\tilde \rho_n =  \left [ \begin{array}{cc}
\langle 1+\tilde\sigma_n^z\rangle / 2 & \langle \tilde\sigma_n^+\rangle \\ \langle \tilde \sigma_n^-\rangle & \langle 1-\tilde \sigma_n^z\rangle  / 2 
\end{array} \right ].
\end{equation}
The off-diagonal matrix can be nonzero if the initial state has a mixture of both even and odd-numbered magnon states and/or if the QDP mixes even and odd states (for example
$\hat n\ne \hat z$ in the Kraus operators above). In all the cases we consider the Hamiltonian dynamics does not mix even and odd magnon states. We will consider three different situations: (A) Initial state has either even-only or odd-only magnon states, QDP does not mix even and odd states ($\hat n=\hat z$), most of the results presented in the next sections will be in this category, where the off-diagonal term above is zero (B) Initital state has both even and odd magnon states (ex. $|000...00\rangle + |100..00\rangle$),  QDP does not mix even and odd states (C) Initial state has even-only magnon states, but the QDP mixes even and odd sectors (for example $\hat n= (\hat x+\hat z)/\sqrt{2}$). In both the  cases of (B) and (C), the
reduced density matrix shown above will have nonzero off-digonal matrix element, thus only the local magnetization $\langle \sigma_n^z\rangle$ cannot determine the local von Neumann
entropy.  

In the situation (A), the Hamiltonian dynamics and the QDP (with $\hat n=\hat z$) both preserve the evenness of the state, i.e. the operators commute with $(-)^{\sum \sigma_l^z}$. Thus,
for an initial state having even-only or odd-only magnon states, the reduced density matrix
will have only diagonal terms in the computational basis, which are completely characterised by  $\langle \sigma_n^z\rangle$. This implies we can construct a QDP detector function
using the expectation value of the diagonal spin operator.
 The local magnetization $\langle \tilde \sigma_n^z\rangle$, calculated from the state $\tilde \rho$ that carries the effect of the QDP occurring at the first spin, will be different from that
 of $\langle \sigma_n^z\rangle$, calculated from the state $\rho$ which is evolved without the QDP, for times larger than the time required for the signal from the QDP location to
 propagate to the $n$'th spin. Consider the detector function defined as,
  \begin{equation}
F_n(t) = \langle\tilde{{\sigma}^{z}_{n}}\rangle_t - \langle{\sigma}^{z}_{n}\rangle_t .
 \end{equation} 
 The detector function defined above uses the contrast between the reduced density matrices from the two states evolved states $\tilde \rho(t)$ and $\rho(t)$. Since, we are only using
 the signal information from a single qubit, from the single-qubit reduced density matrices, this is the simplest detector function,
 There are other detector functions one can construct, for example the relative entropy of the two reduced density matrices, and using two-qubit reduced density matrices, or using even higher multi-qubit information. The above one, the simplest choice, will suffice to detect the QDP  from a distant qubit, as we shall see in the next sections.
 
 The detector function can be rewritten using Eq.7,  we get
 \begin{equation}
F_n(t) = {1\over 2} \langle{\hat n.\vec \sigma_1(t_0)}~\sigma_n^{z}(t)~{\hat n.\vec\sigma_1(t_0)}\rangle
 - {1\over 2} \langle \sigma_n^z(t) \rangle,
 \end{equation}
 where the time-dependent operator is given by ${\sigma}^{z}_{n}(t) = e^{iHt}{\sigma}^{z}_{n}(0)e^{-iHt}$, and the expectation value is taken in the initial state $|\psi(0)\rangle$. Now, the components of the time-evolved operator $\vec \sigma_1(t_0)$ need not commute with $\sigma_n^z(t)$  in general for $t\ne t_0$, unlike
 in the situation considered in Eq.3, due to the dynamics.  This implies that the two terms in the above expression need not be equal, making the detector function nonzero, thus making it possible to detect the QDP signal from other qubits.
It has become possible to detect the occurrence of QDP, due to the dynamics of evolution of the state from $t=t_0$ when the QDP occurs on the first qubit, and at time $t$ when 
 another qubit is measured. The further the measured qubit from the first qubit, the waiting time (or the evolution time) required for detecting QDP is expected to increase, as we
 expect the QDP signal will take time to propagate to a far way qubit. As we shall see in the next sections, the detector function $F_n(t)$ will become nonzero after a waiting time $t_n^*$.
 The waiting time increases with $n$, as further qubits will have to wait longer for the signal to arrive propagating with a finite speed.
 Thus, we can define a signalling speed from the waiting time dependence on the location of the measured qubit. The Schroedinger dynamics, which  itself can induce and redistribute quantum correlations between the various qubits of the system, thus it will influence the efficiency of the detection of QDP.
 The signalling speed will depend in general, as we shall see in the following sections, on the nature of interaction, the strength of interaction, the magnetic field, the distribution of
 quantum correlations present in the initial states.   In most of the cases it is difficult to calculate $F_n(t)$ analytically, and it has to be calculated numerically. We will discuss a few
 well-known models of spin-spin interactions, and a few simple initial states in the context of the QDP signal propagation and detection . In these cases, the detector function can be calculated analytically.
 
 For the situations (B) and (C) where even and odd magnon sectors are mixed due to either the QDP (for $\hat n\ne \hat z$) or initial state being in a mixed sector, we need to investigate
 the dynamics of the off-diagonal matrix element of $\tilde \rho_n$. In analogy with the above detector function, we define an off-diagonal detector function, as
 \begin{equation}
 O_n(t)= \langle \tilde \sigma_n^+(t)\rangle - \langle \sigma_n^+(t)\rangle,
 \end{equation}
 where the two different terms refer to the two states, with and without the QDP occurrence. Now, we have two different time-dependent detector functions. The information carried by
 these two can be combined by looking at the von Neumann entropy of the two reduced density matrices. We can use the excess entropy generated to define a single detector function
 for these situations, we define
 \begin{equation}
 D_n(t)=  - Tr \tilde \rho_n \log \tilde \rho_n  + Tr \rho_n \log \rho_n.
 \end{equation}
 We shall see in the next sections that all these detector functions show similar behaviour, becoming nonzero after a waiting time $t_n^*$ after which the QDP signal propagates to $n$'th
 qubit.
 
 \section{{ Ising Dynamics}}
  The formalism described in the previous section for  the QDP signal detection (with $\hat n =\hat z$ for simplicity) can be applied to one dimensional spin chains of various kind of interactions. Simplest of them is obviously Ising model with nearest neighbour interaction where the Hamiltonian only depends on one kind of Pauli spin matrix. The Hamiltonian of  a
  spin chain of $N$ spins is given by, 
   \begin{equation} 
  H = J\sum_{i=1}{\sigma}^x_{i}{\sigma}^x_{i+1}
  \end{equation} 
The time evolved operator ${\sigma}^{z}_{n}(t)$ is given by,$${\sigma}^{z}_{n}(t)
= e^{it\sigma^x_n(\sigma^x_{n-1}+\sigma^x_{n+1})}{\sigma}^{z}_{n}(0)e^{-it\sigma^x_n(\sigma^x_{n-1}+\sigma^x_{n+1})}.  $$ 
Hence, the operator ${\sigma}^{z}_{n}(t)$ depends only on the operators corresponding to the site indices $(n-1)$ and $(n+1)$. So the operator ${\sigma}^{z}_{1}(t)$ commutes with the operators ${\sigma}^{z}_{n}(t)$ for all $s$ greater than 2. In this case, from Eq.10 we see that the two terms become equal making the detector function to vainish, and the dynamics becomes trivial. Fig.2(a) shows the plot of the time evolution of the detector function as a function of time for the second, the third and the fourth qubits. The detector function is nonzero only for
the first neighbour of the site at which the QDP occurs, and displays a periodic behaviour due to the nature of the dynamics. This would imply that it is not possible to detect the signal form any other site except the second one as shown in the figure, and thus the speed of the signal propagation cannot be defined. This is true for any initial state of the system irrespective of whether the initial state is entangled or not. These features can be attributed to the simplest dynamics that we are considering with only nearest neighbour interactions.
 \begin{figure*}
\begin{tabular}{cc}
\includegraphics[width=0.5\textwidth]{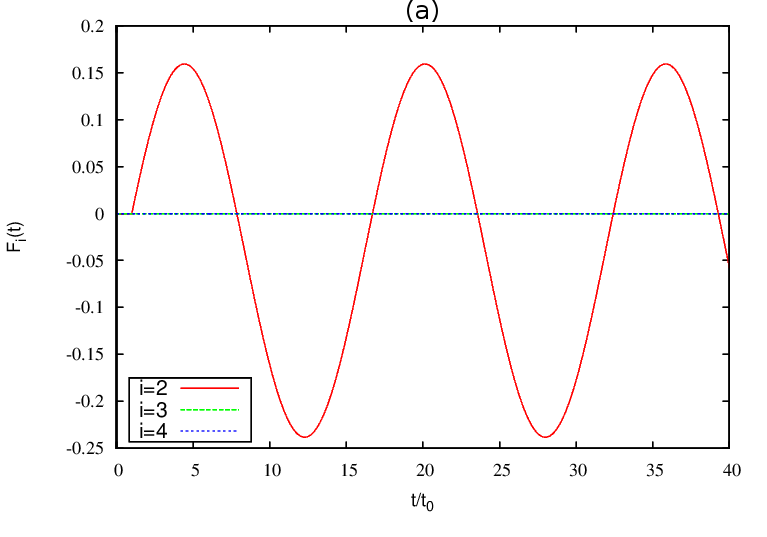}&
\includegraphics[width=0.5\textwidth]{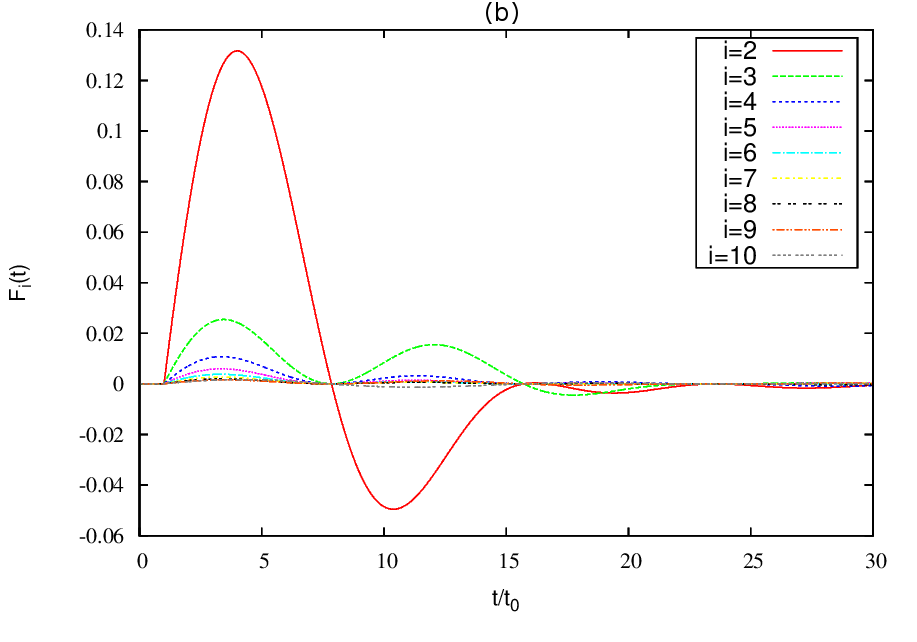} \\
\includegraphics[width=0.50\textwidth]{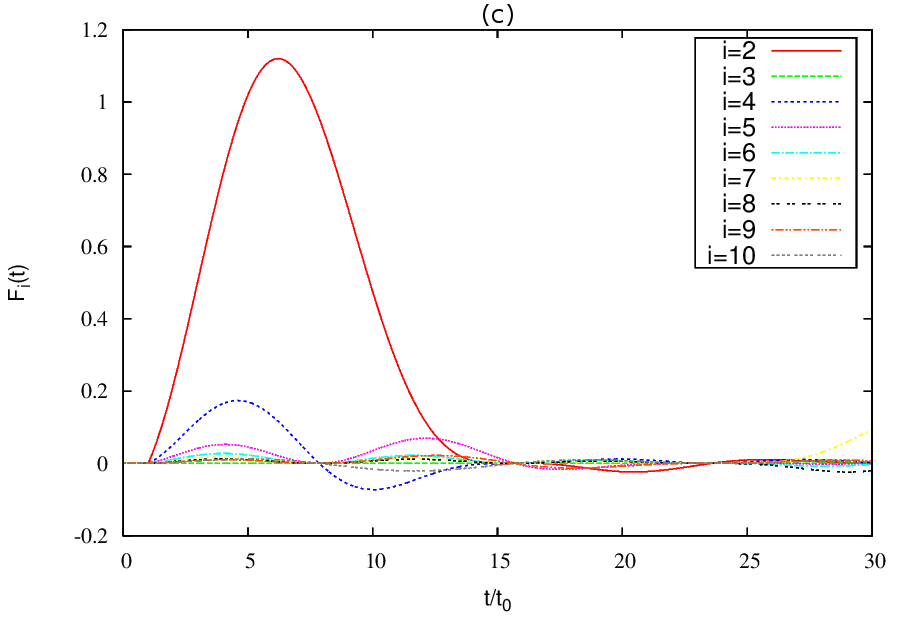}&
\includegraphics[width=0.50\textwidth]{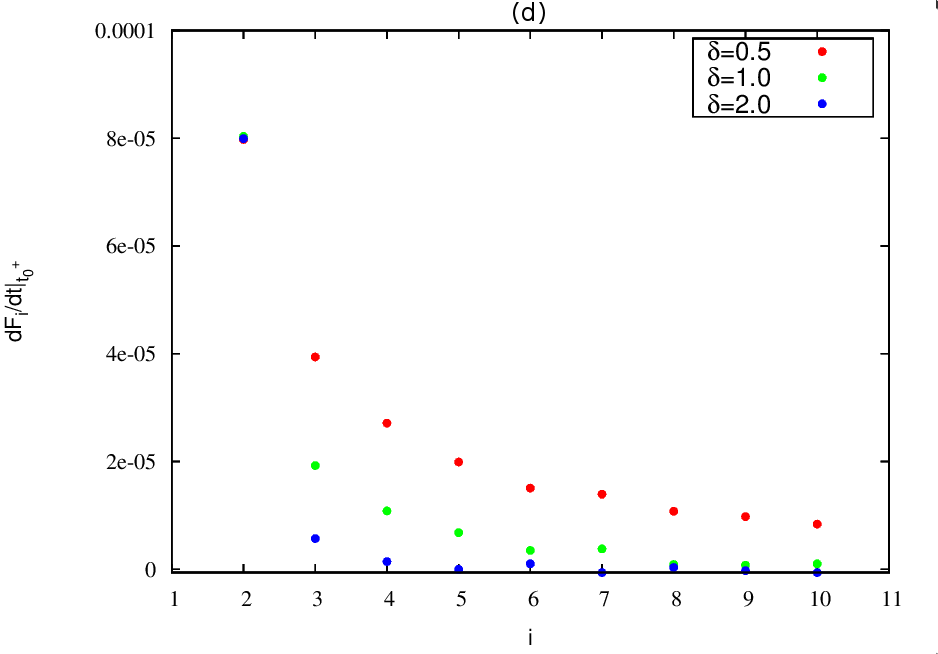}
\end{tabular}
\caption
{\small The detector function $F_n(t) $ is plotted as a function of time for Ising dynamics with open boundary conditions and the QDP with $\hat n=\hat z$: (a) For the initial state $|00...0\rangle$
using the nearest-neighbour Ising dynamics. The QDP signal does not reach further than the second site in this case.
(b) For the initial state $|00...0\rangle$ using the long-ranged Ising dynamics (c) For the initial state $(|100...0\rangle+|0010..0\rangle)/\sqrt{2}$ for the long-ranged Ising model. For the long-ranged Ising case, the signal reaches all other sites instantaneously at $t=t_0$. (d) The slope of the detector function  $t_0(dF/dt)_{t=t_{0^+}}$  is plotted against the site index. The derivative decreases as we go further from the first site, because the strength of interaction decreases with the distance between the spins as a power law of the distance between them as depicted in Eq.14.
The total number of sites $N=10$ in all the cases, and  $\delta=1$ for (b) and (c). }
 \end{figure*}

To see a non trivial behaviour in the dynamics, we consider the Ising model with spatially long-ranged interactions among the spins. Here each of the spins interacts with all other spins and the interaction strength follows a inverse power law determined by the parameter $\delta$. The Hamiltonian is given by 
 \begin{equation} 
 H = \sum_{<i,k>}\frac{J}{|i-k|^{\delta}}{\sigma}^x_{i}{\sigma}^x_{k},
 \end{equation} where the sum is over all pairs. Such long-ranged interactions,  which can be realized in ion traps by controlling the intensity and polarization of laser fields, can be
 used to study quantum phase transitions in quantum spins\cite{Porras}.
Now, the time evolved operator ${\sigma}^{z}_{n}(t)$ depends on the operators corresponding to all sites. The expression for ${\sigma}^{z}_{n}(t)$ is given as,
  \begin{eqnarray}
  {\sigma}^{z}_{n}(t)
  & = &\prod_{k\neq n}e^{it{\sigma}^{x}_{n}{\sigma}^{z}_{k}\frac{J}{|n-k|^{\delta}}}{\sigma}^{z}_{n}(0)e^{-it{\sigma}^{x}_{n}{\sigma}^{z}_{k}\frac{J}{|n-k|^{\delta}}},\nonumber
  \\&=&(A_n^2 -B_n^2){\sigma}^{z}_{n}(0) - 2A_nB_n{\sigma}^{y}_{n}(0).
  \end{eqnarray}
 Here we introduced the operators $ A_n$ and $B_n$, which are given as,
  \begin{eqnarray}
  A_n &\equiv& {\rm Re}  \prod_{k \neq n} \left [ \cos(\frac{Jt}{|n-k|^{\delta}})+ i \sigma_k^x \sin(\frac{Jt}{|n-k|^{\delta}})\right ], \nonumber \\
  B_n &\equiv& {\rm Im}  \prod_{k \neq n} \left [\cos(\frac{Jt}{|n-k|^{\delta}})+ i \sigma_k^x \sin(\frac{Jt}{|n-k|^{\delta}})\right ].
   \end{eqnarray}
  The expectation value of the operator ${\sigma}^{z}_{n}(t)$ can be calculated for the initial state $\rho(0) = |00..0\rangle\langle00..0|$ as,
  \begin{eqnarray}
  \langle{\sigma}^{z}_{n}\rangle_t 
  &=& Tr[((A_n^2 -B_n^2){\sigma}^{z}_{n}(0) - 2A_nB_n{\sigma}^{y}_{n}(0))\rho(0)], \nonumber \\ &=&  \prod_{k \neq n} cos \left(\frac{2J}{|n-k|^{\delta}}t\right).
  \end{eqnarray}
However, in this case the analytical calculation for $F_n(t)$ is difficult as it contains three time dependent operators. We use a computational method, as discussed below.

For this Hamiltonian the energy eigenstates are the direct product states of $\sigma^x$ eigenstates,  $|s_1 s_2..s_N\rangle$, and the eigenvalues are given by
$$E(s_1 s_2..s_N) = \sum_{\langle i,k\rangle}\frac{J}{|i-k|^{\delta}}(-1)^{s_i+s_k},$$
where $s_i$ is the eigenvalue of $\sigma^x_i$.
Let $|\Psi(0)\rangle$ be the initial state of the system, which can be expanded in the eigen basis ${|a_i\rangle}$ (which happens to be eigen basis of ${\sigma}^{x}$ in this case) of the Hamiltonian as $|\Psi(0)\rangle = \sum_{i} c_i|a_i\rangle$. The state at any later time is given by $|\Psi(t)\rangle = \sum_{i} c_i e^{-iE_it}|a_i\rangle$. The density matrix at time $t=t_0$ becomes ${\rho}(t_0) = \sum_{i,j} c^*_jc_i e^{i(E_j-E_i)t_0}|a_i\rangle\langle a_j|$.
 Just after the QDP (with $\hat n=\hat z$)  occurs at the first site the state becomes, at $t=t_0^+$,
 \begin{equation} \tilde{{\rho}}(t_{0}^{+}) = \frac{1}{2}\sum_{i,j} c^*_jc_i e^{i(E_j-E_i)t_0} \{~|a_i\rangle\langle a_j| + 
{\sigma}^{z}_1 |a_i\rangle\langle a_j|{\sigma}^{z}_1~\}
 \end{equation} 
 Evolving the state from $t_0$ to $t$ we get the state $\tilde{\rho}(t)$ that has two terms as shown in Eq.7.
 Now, we can numerically compute the inner product of the initial state $|\Psi(0)\rangle$ with the eigenstates of the Hamiltonian to find the the state $|\tilde{\Psi}(t)\rangle$ and hence the detector function $F_n(t)$.
 
Now, we discuss below the results for specific initial states. For the initial state $|00...0\rangle$ we have the detector function $F_n(t)$ non zero for all sites for any time $t > t_0$ as depicted in Fig 2(b). So, the observer can detect the signal from any site without any delay. But this is not true for entangled states. Let us consider the state $(|100..0\rangle+|010..0\rangle)/\sqrt{2}$, where the first spin is maximally entangled with the second one; in this case we see the detector function $F_n(t)$ is zero only for the second site. The general argument is given as follows. If first site is maximally entangled with $m^{th}$ site, the value of $F_m(t)$ is zero for any time. This can be explained as follows.\\
\\ For example, let the initial state be $|\Psi(0)\rangle = (|0_10_m\rangle + |1_11_m\rangle)\otimes|00...0\rangle/\sqrt{2}$. Here the  first and the $m^{th}$ site is maximally entangled. The terms like $\langle a_j|{\sigma}^{z}_{m}|a_i\rangle$ will be non zero only if the  $m^{th}$ site configuration  of the states $|a_i\rangle$ and $|a_j\rangle$ (see Eq. 16) is different and rest are same. However, both the expressions of  $F_n(t)$ and $\langle{\sigma}^{z}_{n}\rangle_t$ contain the coefficients $c_i$ and $c_j$ which are nothing but the inner product of eigenstates of $\sigma^x$ basis with the initial state of the system. Either of these two states have odd number of ($|S^x,-\rangle$) spins. Since the initial state has a definite parity either $c_i$ or $c_j$ will be zero. This gives $F_m(t) = 0$ for all values of t. Similar argument can also be given for the initial state $|\Psi(0)\rangle = (|1_10_m\rangle + |0_11_m\rangle)\otimes|00...0\rangle/\sqrt{2}$. Fig 2(c) shows that $F_3(t)$ is zero for the third site when the initial state is $(|100...0\rangle+|0010..0\rangle)/\sqrt{2}$. \\ 
\\ Extending the argument for GHZ like states we can say all the states are maximally entangled with each other. Hence the detector function  $F_m(t) = 0$ for all values of $m$ for all time.\\
\\We have seen that for nearest-neighbour interactions,  only $F_2(t)$ is non zero and $F_n(t)$ vanishes for other sites. In the case of lang-ranged interactions, the first site interacts with all other sites with different interaction  strengths determined by the Hamiltonian of the system. Now, the detector function $F_n(t)$ is non zero for all sites for time $t>t_0$. As the distance between the sites increase the interaction strength decreases by power law given in Eq(12). The slope function $(dF/dt)|_{t_0}$ is plotted against the site index in Fig 2(d), exhibits this  same trend. The speed of the signal propagation cannot  be defined for the Ising dynamics because the signal does not propagate beyond the second site in the case of nearest-neighbour interactions, and the signal reaches all the sites without any delay for the case of long-ranged interactions.   

\section{{Anisotropic Heisenberg model}}
One of the first exactly solvable but non trivial models of quantum mechanics is a one dimensional chain of spins interacting with their nearest neighbour Heisenberg exchange interaction, known as the Heisenberg model\cite{ref7}. Three kinds of Pauli spin matrices in the Hamiltonian indicates interactions in all three spin dimensions. The anisotropic Heisenberg model Hamiltonian for a spin chain of $N$ spins is given by,
 \begin{equation}
H = \sum_{i}J({\sigma}^x_{i}{\sigma}^x_{i+1}+{\sigma}^y_{i}{\sigma}^y_{i+1})+J_z{\sigma}^y_{i}{\sigma}^y_{i+1}.
 \end{equation}
 For the case of $J_z=0$, this Hamiltonian can be mapped to a free fermion model. For $J_z>0$, the model exhibits antiferromagnetic behaviour, and it is ferromagnetic for $J_z<0$. The
 model exhibits a Kosterlitz-Thouless-type quantum critical point for the isotropic case of $J_z=J$. Since the total z-component of the spin commutes with the Hamiltonian,
the eigenstates of this Hamiltonian have a definite number spins in state $|1\rangle$ (down
spin). An eigenstate with $l$ down spins, a $l$-magnon state, can be written as,
$$ |\psi\rangle = \sum_{x_1,x_2..x_l}\psi_{x_1,x_2..x_l}|x_1,x_2..x_l\rangle,$$
where the basis state is labeled by the locations of the $l$ down spins. The eigenfunction $\psi_{x_1,x_2..x+l}$ is given by the Bethe ansatz\cite{ref19}, which will be labeled by the the set of momenta
$k_i$ of the down spins, that are determined by solving $l$ algebraic Bethe ansatz equations, with periodic boundary conditions. The interaction strength $J$ determines the hopping of
the down spins to nearby sites, whereas the interaction of the two down spins is determined by $J_z$. The one-magnon eigen energies are independent of $J_z$ as the states carry
only one down spins. For $l>1$, the eigenstates include both scattering states of one magnon, and  of the down spins. It is straightforward to see the states $|00...0\rangle$ and $|11...1\rangle$ are eigenstates of the Hamiltonian, where $|0\rangle$, and $|1\rangle$ are eigenstates of $\sigma_i^z$.  We have seen that if the initial state is this eigenstate of the Hamiltonian no dynamics will be observed. The dynamics of the magnon bound states using a quench have been studied\cite{refa1}. We will investigate a linear combination of zero-magnon, one-magnon and two-magnon states that are not eigenstates of the Hamiltonian, using the QDP dynamics below. For the most part we consider $\hat n=\hat z$, and will
consider the case of $\hat n\ne \hat z$ towards the end of the section to illustrate the effect of mixing the even and odd sector states.

Let us consider a general initial state, to observe any non trivial dynamics, given by
 \begin{equation}
|\Psi(0)\rangle = \alpha|x\equiv\{x_1x_2..x_l\}\rangle+\beta|y\equiv \{y_1y_2..y_m\}\rangle,
 \end{equation}
 where $x_i(y_i)$ denotes the location of the $i'$th down spin. That is, this state is superposition of the states with $l$ and $m$ number of down spins; $(x_1<x_2<...)$ are the co-ordinates of the sites with down spins in the first part, and similarly for the second part. Such a state can be written as a linear combination of momentum eigenstates $|k\equiv\{k_1,k_2..k_l\}\rangle$,
 and, $|q\equiv \{q_1,q_2..q_,\}\rangle$ of the Hamiltonian, with eigenvalues $\epsilon(k),\epsilon(q)$. Hence,
  \begin{equation}
|\Psi(0)\rangle = \alpha \sum_{k_1..k_l}\psi^{k}_{x}|k\rangle +\beta \sum_{q_1..q_m}\psi^{q}_{y}|q\rangle,
 \end{equation}
 where the wave functions denote $\psi^k_x\equiv \langle k_1..k_l|x_1..x_l\rangle$, and $\psi^q_y\equiv \langle q_1..q_m|y_1..y_m\rangle$. Now, the time evolution of the state is straightforward, the state after a time $t$ becomes,
   \begin{equation}
|\Psi(t)\rangle = \alpha \sum_{x'_1..x'_l}G^{x'}_{x}(t)|x'_1..x'_l\rangle +\beta \sum_{y'_1..y'_m}G^{y'}_{y}(t)|y'_1..y'_m\rangle
 \end{equation}
 where the time-dependent function $G$ is given in terms of the wave functions defined above as,
   \begin{equation} 
 G^{x'_1..x'_{l}}_{x_1..x_{l}}(t)=\sum_{k_1..k_{l}}\psi^{k_1..k_{l}}_{x_1..x_{l}}  \psi^{* k_1..k_{l}}_{x'_1..x'_{l}} e^{-it\epsilon(k_1..k_{l})}
  \end{equation}
Expectation value of $\sigma_n^z$ for the $n^{th}$ site is then given by,
\begin{equation}\langle\sigma_n^z\rangle_t =1-2|\alpha|^2 {\sum_{x_1'..x_l'}}^\prime |G^{x'_1..n..x'_l}_{x_1,x_2..x_l}|^2 -2 |\beta|^2 {\sum_{y_1'..y_l'}}^\prime |G^{y'_1..n..y'_l}_{y_1,y_2..y_m}|^2,\end{equation}
where the prime over the sums in the above indicates that there is one less free variable to be summed, that is $l-1~(m-1)$ number of $x_i (y_i)$ variables in the first (second) sum.
 Just after the system undergoes a QDP (with $\hat n =\hat z$) on the first site the state at $t = t_0$ is given by,
  \begin{equation}
\tilde{{\rho}}(t_{0^+})=|\tilde \Phi_+(t_0)\rangle \langle \tilde \Phi_+(t_0)|+|\tilde \Phi_-(t_0)\rangle \langle \tilde\Phi_-(t_0)|,
    \end{equation}
where, $|\tilde \Phi_{\pm}(t_0)\rangle\equiv {1\pm \sigma_1^z\over 2}|\psi(t_0)$, we have 
   \begin{eqnarray}
|\tilde \Phi_+(t_0)\rangle &=&\alpha \sum_{x'_1\neq 1,x'_2..x'_l
}G^{x'}_{x}(t_0)|x'\rangle +\beta \sum_{y'_1\neq 1,y'_2..y'_m
}G^{y'}_{y}(t_0)|y'\rangle, \nonumber\\
|\tilde \Phi_-(t_0)\rangle &=&\alpha \sum_{x'_1= 1,x'_2..x'_l
}G^{x'}_{x}(t_0)|x'\rangle +\beta \sum_{y'_1=1,y'_2..y'_m
}G^{y'}_{y}(t_0)|y'\rangle
   \end{eqnarray}
Further evolution of the system for a time $(t-t_0)$ yields the state at time $t$ ,
 \begin{equation}
\tilde{{\rho}}(t)=|\tilde\Phi_+(t)\rangle \langle \tilde\Phi_+(t)|+|\tilde\Phi_-(t)\rangle \langle \tilde\Phi_-(t)|,
 \end{equation}
 where,
    \begin{eqnarray}
 |\tilde \Phi_+(t)\rangle&=&\alpha \sum_{x'_1..x'_l}H^{x'}_{x}(t,t_0)|x'\rangle +\beta \sum_{y'_1..y'_m}H^{y'}_{y}(t,t_0)|y'\rangle,\nonumber\\
|\tilde\Phi_-(t)\rangle&=&\alpha \sum_{x'_1..x'_l}K^{x'}_{x}(t,t_0)|x'\rangle +\beta \sum_{y'_1..y'_m}K^{y'}_{y}(t,t_0)|y'\rangle.
   \end{eqnarray}
Here, the new time-dependent wave functions are given by,
\begin{eqnarray}
H^{x'}_{x}(t,t_0)= \sum_{x''_1\neq 1,x''_2..x''_l
} G^{x''}_x(t_0) G^{x'}_{x''}(t-t_0),\nonumber\\
K^{x'}_{x}(t,t_0)= \sum_{x''_1= 1,x''_2..x''_l
} G^{x''}_x(t_0) G^{x'}_{x''}(t-t_0).
\end{eqnarray}
\begin{figure*}\begin{tabular}{cc}

       \includegraphics[width=0.50\textwidth]{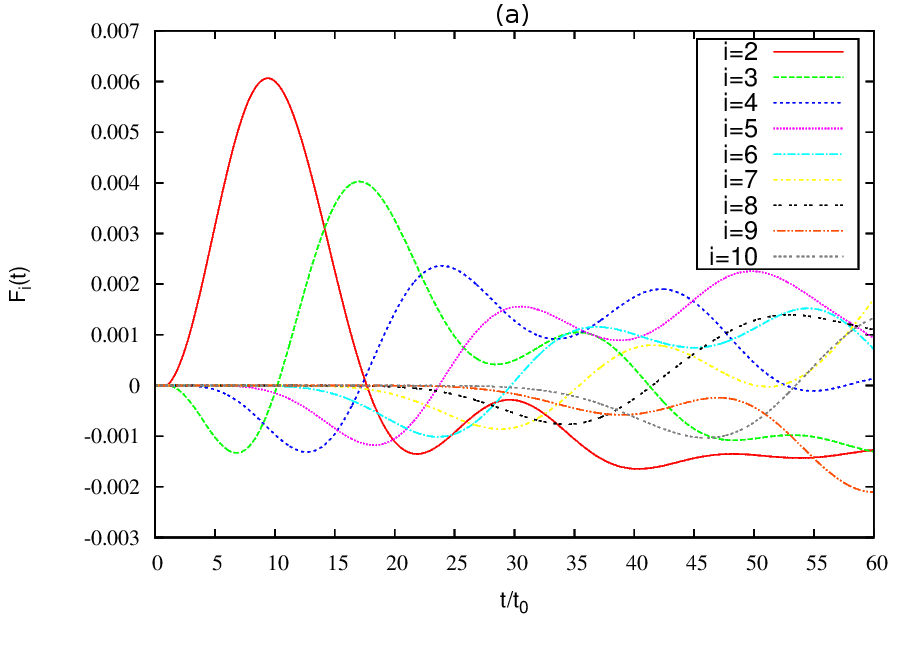}&
	 \includegraphics[width=0.50\textwidth]{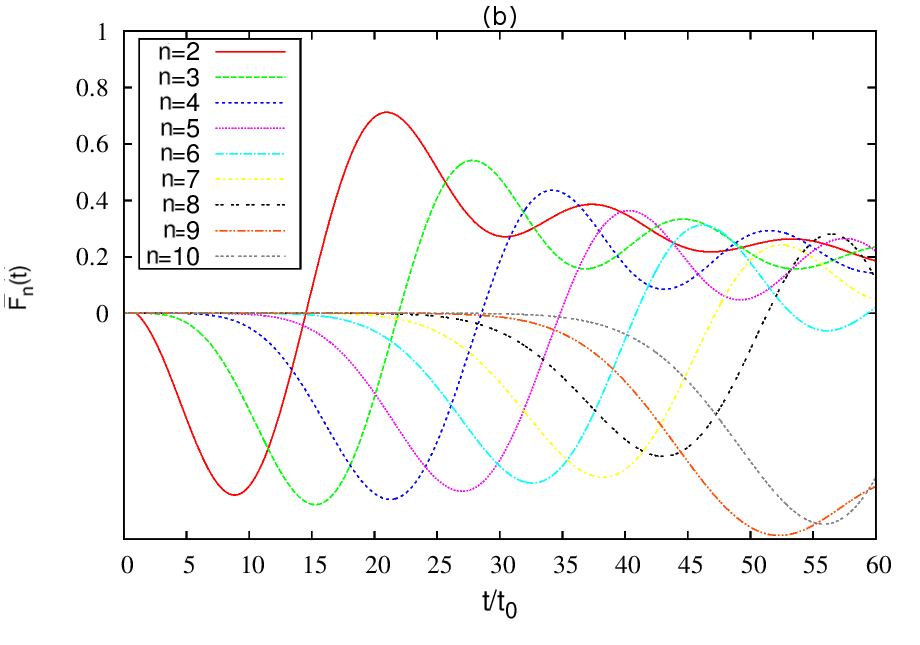}\\
    \includegraphics[width=0.50\textwidth]{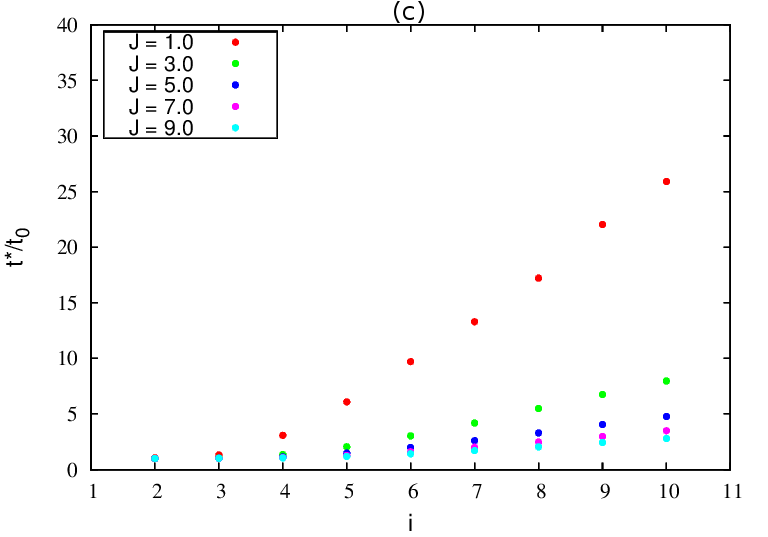}&
       \includegraphics[width=0.50\textwidth]{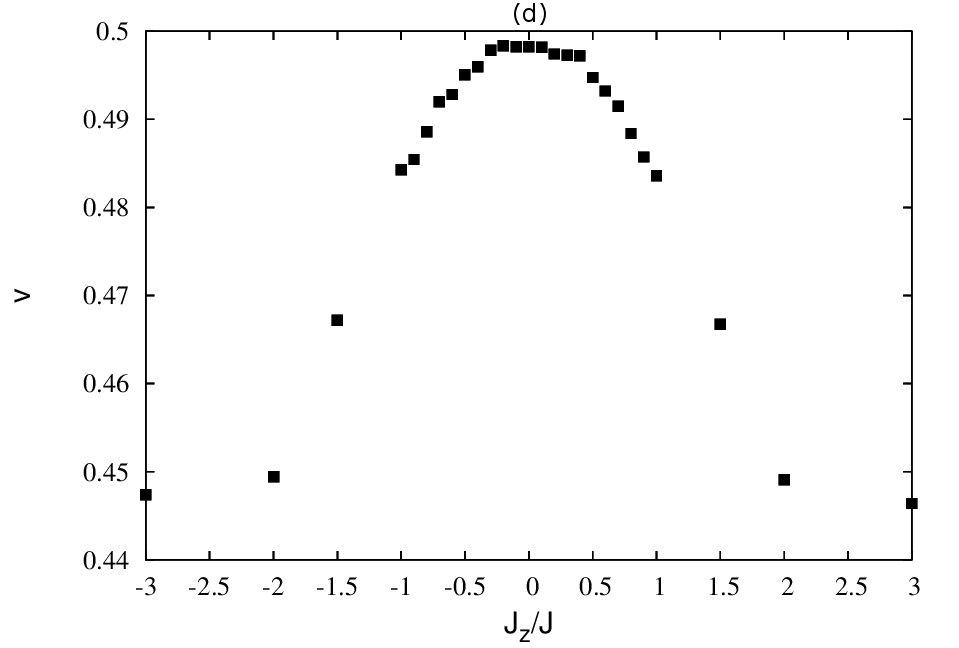}       
       \end{tabular}
\caption{\label{fig:fig_2}{\small 
The detector function $F_n(t) $ is plotted as a function of time for the Heisenberg dynamics with open boundary conditions and the QDP with $\hat n=\hat z$: (a) For the even-magnon initial state $|000...0\rangle+|110..0\rangle/\sqrt{2}$ for the isotropic Heisenberg model with interaction strength $J=1$  (b) For the one-magnon initial state 
$|010...0\rangle+|100..0\rangle/\sqrt{2}$ 
for isotropic Heisenberg model with interaction strength $J=1.0$ with periodic boundary conditions (c) $t^*/t_0$ is  plotted with $n$  for different values of $J$ for the even-magnon initial state $|000...0\rangle+|110..0\rangle/\sqrt{2}$; $t^*$ is time when $|F_n(t)|$ just exceeds a small positive number $\epsilon$ (taken to be $10^{-5}$) for the first time.The curves attain roughly a constant slope for different values of $J$. The speed of propagation is determined by the inverse of the slope. (d) The speed is plotted against $J_z/J$ for the anisotropic model, where the reference speed $v_0=10 |J| a / \hbar$  in terms of $a$, the nearest-neighbour separation of the qubits.}}
\end{figure*}
Using the above functions, a simpler form for the QDP detector function for the $n'$th qubit is given as,
\begin{eqnarray}
F_n&(t)=2|\alpha|^2 \sum_{x'}|G_{x_1,x_2..x_l}^{x'_1..n..x'_l}|^2 -|H_{x_1,x_2..x_l}^{x'_1..n..x'_l}|^2-|K_{x_1,x_2..x_l}^{x'_1..n..x'_l}|^2 \nonumber\\
&+2|\beta|^2 \sum_{y'}|G_{y_1,y_2..x_m}^{y'_1..n..y'_m}|^2 - |H_{y_1,y_2..y_m}^{y'_1..n..y'_m}|^2-|K_{y_1,y_2..x_m}^{y'_1..n..y'_m}|^2.
\end{eqnarray} 
 In the above, the sums are over the variable sets $x'$ and $y'$ that include the position $n$ as one of the elements, as explicitly shown in the superscripts of the time-dependent functions
 $G, H,$ and $K$.  
 
Let us first consider the state $(|100...0\rangle+|010...0\rangle)/\sqrt{2}$. Since this is a one magnon state ($l=m=1$) equation(19) simplifies to
  \begin{equation}
F_n(t)=\sum_{x=1,2}|G^n_x(t)|^2 - |H^n_x(t,t_0)|^2 -|K^n_x(t,t_0)|^2.
 \end{equation}
 For this case, the one-magnon eigenfunction is given by, $\psi(x)= e^{ikx}/\sqrt{N}$, and the eigenvalue is given by $\epsilon_k= 4J \cos{k}$, apart from a constant
 that depends on $J^z$, the diagonal interaction term in the Hamiltonian\cite{ref19}.
 We can use the time scaled by $\hbar/4J$, the natural time unit from the spin exchange interaction. 
 In the macroscopic limit, $N\rightarrow\infty$, the sum over the momentum in Eq.13 can be converted
 into an integral, thus, we get
  \begin{equation}
  G^n_x(t)={1\over 2 \pi}\int^{2\pi}_{0} e^{-it\cos{k}-ik(x-n)}dk= J_{x-n}(t)i^{x-n}. \end{equation}
Now, using this we can determine the other two time-dependent functions that are needed in $F_n(t)$, we have  
\begin{eqnarray}
H_1^n&=i&^{1-n}(J_{1-n}(t)-J_0(t_0)J_{1-n}(t-t_0)), ~H_2^n=i^{2-n}J_{2-n}(t-t_0),\nonumber \\
K_1^n&=&i^{1-n}J_0(t_0)J_{1-n}(t-t_0), ~K_2^n=i^{2-n}J_1(t_0)J_{1-n}(t-t_0).
\end{eqnarray}
Fig 3(b) shows the detector function $F(t)$ as a function of time for initial state $(|100...0\rangle+|010..0\rangle)/\sqrt{2}$. Unlike Ising dynamics here we can see that the detector function is zero at time $t =t_0$ for all sites and farther the site from the first one, later it becomes non zero. Hence, it is possible to define a speed in this case. To calculate the speed numerically we define a time instant $t^*$ when  the detector function $|F_n(t)|$ just exceeds a small positive number $\epsilon$  for the first time. It has been seen from figure3(c) that the plot of the 
quantity $t^*$ versus site indices becomes linear,  $t^*/t_0 \approx  {w} n$, the slope ${ w}$ determines the speed of signal propagation. Let $a$ denote the physical separation between neighbouring qubits,  the distance from the QDP site is given by $r=na$. Thus, we can define the speed of the signal propagation through $ r = v t^*$, we have $v=v_0/{  w}$, where the reference speed is given by $v_0=a/t_0=10|J|a/\hbar$ (as we have used $J t_0/ \hbar=0.1$ for the numerical plots) in terms of the coupling strength and the separation between the qubits.  Using a typical parameter value set pertaining to spin chains, $J=10 meV$ and $a=0.1 nM$, we get an
estimate of $v_0 \sim 10^4 M/sec$. We can see from Fig.3d that the speed of propagation $v$ is of the same order as $v_0$. 
 
 We now consider the initial state $(|00...0\rangle+|110...0\rangle)/\sqrt{2}$, which is a linear combination of the two-magnon eigenstates with $l=0, m=2$. The first part of the state, with $l=0$ is an eigenstate of the Hamiltonian, which does not change in time till $t_0$. The quantum channel action on the first site does not alter the state either. Thus it does not contribute to
 the dynamics altogether. The second part of the state, with $m=2$ a two-magnon state,  does have very complicated dynamics, with both two one-magnon scattering states and two-magnon bound states contributing to the dynamics. The eigenfunctions and eigenvalues can be obtained from Bethe ansatz\cite{ref19}, however, the time-dependent functions are quite difficult to
 calculate. The detector function is given by,
  \begin{equation}
F_n(t)=\sum_{y_1}|G_{1,2}^{y_1,n}(t)|^2 - |H_{1,2}^{y_1,n}(t,t_0)|^2- |K_{1,2}^{y_1,n}(t,t_0)|^2,
 \end{equation}
 where the sum is over all site indices except $n$, and the superscript in all the functions is an ordered set.
 Unlike the previous case of one-magnon states, here both bound and scattering states are possible. In the case of a infinitely long chain it can be shown that the  quasi-momenta become continuous set of numbers from $0$ to $2\pi$ and $(-\infty+i/2)$ to $(-\infty+i/2)$ for scattering and bound states respectively. However, any closed form of $F_n(t)$ is difficult to find. In this case we have calculated the quantity $F_n(t)$ by calculating the time-dependent functions $G, H,$ and $K$  numerically. It is seen that the result exactly matches with the result from  exact numerical diagonalization. Fig 3(a) and Fig 3(b) show F(t) for the two initial states $(|100...0\rangle+|010..0\rangle)/\sqrt{2}$ and $(|00...0\rangle+|110..0\rangle)/\sqrt{2}$ respectively.\\
 
Fig 3(a) shows the detector function $F(t)$ as a function of time for initial state $(|000...0\rangle+|110..0\rangle)/\sqrt{2}$. It is clear from the figure that the speed does not depend on the initial states or entanglement of the initial state, though the dynamics varies.  We expect that the speed should increase with the interaction strength parameter $J$  in the case of isotropic Heisenberg model . Fig 3(c) shows that the time $t^*$ decreases as the value of $J$ increases for a fixed site. If we define the speed as $v = \frac{\Delta s}{\Delta (t^*/t_0)}$ it can be shown that the speed is linearly proportional to $J$.  
 \begin{figure*}\begin{tabular}{ccc}
\includegraphics[width=0.3\textwidth]{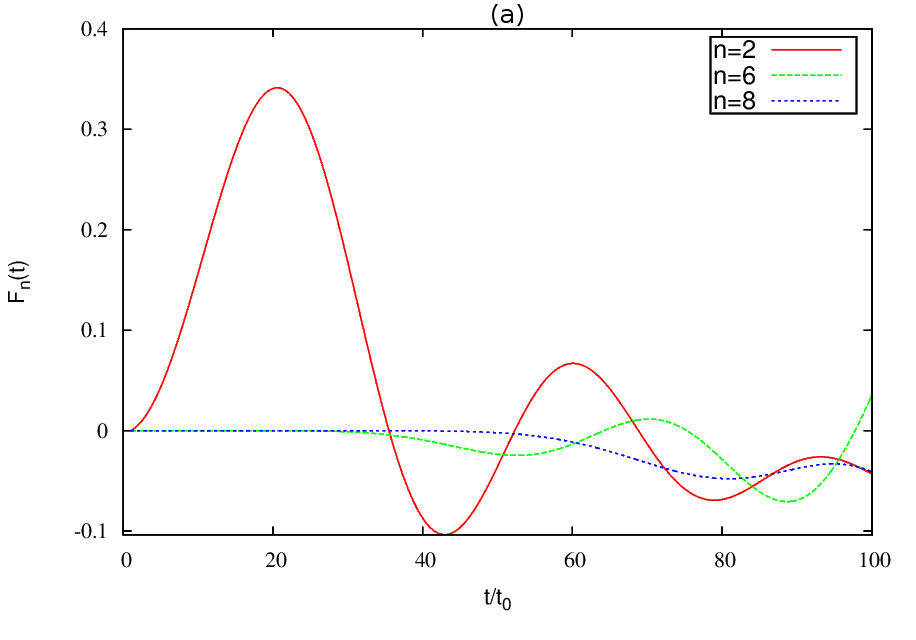}&
	 \includegraphics[width=0.3\textwidth]{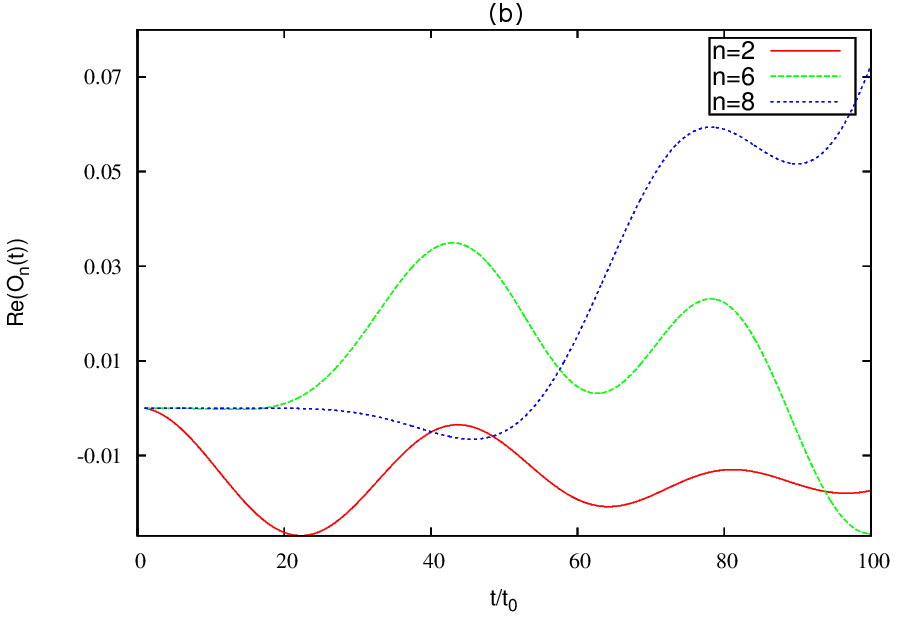}&
    \includegraphics[width=0.3\textwidth]{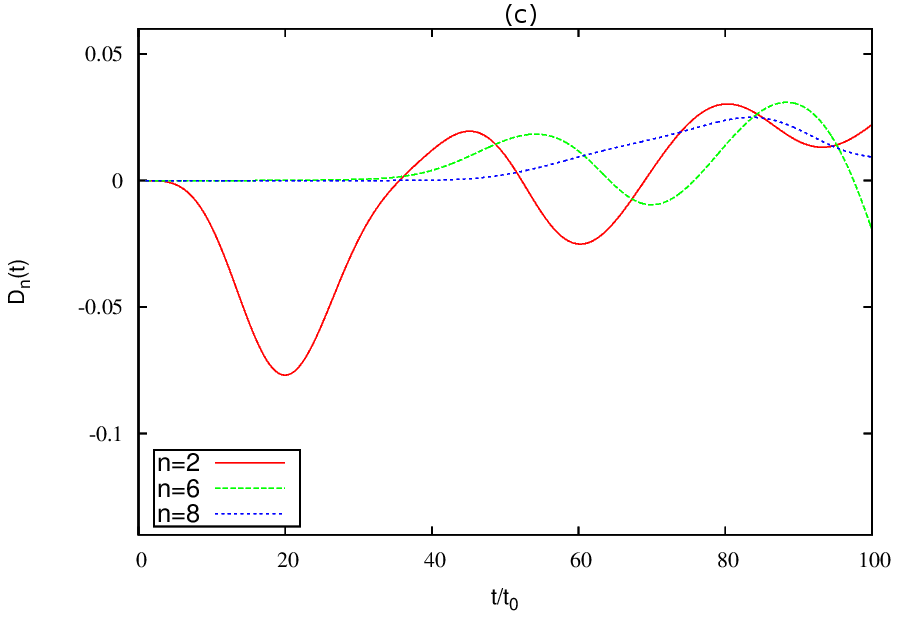}\\
       \end{tabular}
 	\caption{\label{fig:fig_4p}{\small The diagonal and off-diagonal detector functions $F_n$ and $O_n$, and the excess entropy function $D_n$ are plotted as functions of $t/t_0$ respectively for the Heisenberg dynamics with open boundary conditions, and the QDP with $\hat n=  \cos{\theta} \hat z  + \sin{\theta} \hat x$ for a few values of $n$. The interaction strengths are $J_x=1.0, J_y=1.0$ and $J_z=0.5$, the initial state is $|000...0\rangle+|110..0\rangle/\sqrt{2}$, and $\theta = \pi/3$. These functions show similar behaviour as that of $F_n$ for
	the QDP with $\hat n=\hat z$ shown in Fig.3.	 }}
\end{figure*}
 In the case of anisotropic Heisenberg model,  if $|J_z|>> |J|$ the dynamics become trivial and no speed will be observed. So we expect maximum speed when the quantity $J_z$ is zero, which is borne out in Fig 3(d). It can be seen from the figure that the speed of detection itself has a small variation with the parameter $J_z/J$, varying in a range of about ten per cent from
the maximum value. This is, as argued in Section II, due to the fact that QDP has been affected only on one spin at an epoch time for the state $\tilde \rho(t)$, and the subsequent unitary evolution is same as the evolution of the state $\rho(t)$ with no QDP. Both these states would differ only to this extent. We need to study the contrast in the spreading of quantum correlations between the two states, which involves higher marginals, ex. two-qubit reduced density matrices that carry information about pairwise entanglement. The method developed in
this section can be used significantly in studying the effect of multi-party correlations and mutual information.

Let us consider the initial state as $|00..0\rangle$ or $|11..1\rangle$. These states are the eigenstates of the Hamiltonian with same eigenvalue. We can see from Eq(17) that $\tilde{\rho}(t) = \rho(0)$. This implies no dynamics will be seen for these states. Next, we consider the GHZ like state $(|00...0\rangle+|11..1\rangle)/\sqrt{2}$  where each of the site is globally entangled with all other sites. In this case $\langle \sigma^z_n\rangle_t = Tr (\sigma^z_n(t) \rho(0)) = 1$. Also from Eq(17) it can be shown that $\langle \tilde{\sigma^z_n}\rangle_t = Tr (\sigma^z_n(t) \tilde{\rho}(0)) = 1$. Hence the detector function will be zero for this state.

Now, we consider a QDP with $\hat n = \cos{\theta} \hat z + \sin{\theta} \hat x$ and an initial state $|\psi(0)\rangle=|000..0\rangle + |110...0\rangle/\sqrt{2}$. As argued in Sec. II,  the reduced density matrix
$\tilde \rho_n$ (see Eq. 8) will have a nonzero off-diagonal matrix element. The Hamiltonian unitary dynamics does not mix even and odd magnon states, but the QDP explicitly mixes
them. The zero magnon state $|000..0\rangle$ does not evolve under the unitary evolution, but it becomes a mixture of zero and one magnon states after the QDP occurs. Similarly,
the two magnon state will yield a mixture of one, two and three magnon states. Unlike in the previous examples, we have four different magnon sectors contributing for the state
$\tilde \rho(t)$, with both even and odd sectors. We need to investigate detector functions $F_n(t)$ and $O_n(t)$ corresponding to the diagonal and off-diagonal matrix elements,
as discussed in Sec.II. These detector functions have more complicated expressions in this case as there are four different magnon sector contributing to the dynamics for $t>t_0$.

The state $\tilde \rho(t_0)$ just after the QDP occurs is as show in Eq.26, is composed into two pure states, but the two pure states in this case are defined as $|\tilde \Phi_{\pm}(t_0)\rangle
\equiv {1+\cos{\theta}\sigma_1^z+\sin{\theta}\sigma_1^x\over 2}|\psi(t_0)$. The operator $\sigma_1^x$ will create one and three magnon states from zero and two magnon components
present in the initial state. Further evolving the state to a time, the two pure states are given by (analogous to Eq.29),
\begin{eqnarray} 
{\hskip -1.2 cm} |\tilde \Phi_{\pm}(t)\rangle  &= & e^{-iE_0t} \cos^2{\theta\over2} ~|00...0\rangle\pm {\sin{\theta}\over2} \sum_{x'_1}~ [ e^{-iE_0t_0}G^{x'_1}_{1}(t-t_0)+L^{x'_1}_{1,2}(t,t_0) ]~ |x'_1\rangle \nonumber \\ &\pm& 
 \sum_{x'_1,x'_2} [ \cos^2{\theta\over 2}  H^{x'_1,x'_2}_{1,2}(t,t_0)  +{1\mp\sin{\theta}\over 2}K^{x'_1,x'_2}_{1,2}(t,t_0) ]|x'_1,x'_2\rangle \nonumber\\ 
& \pm & {\sin{\theta}\over 2}\sum_{x'_1,x'_2,x'_3}   L^{x'_1,x'_2,x'_3}_{1,2}(t,t_0)|x'_1,x'_2,x'_3\rangle.
 \end{eqnarray}
We have defined new propagator functions connecting the even and odd sectors, defined in terms of the previously defined $G$ functions, as
\begin{eqnarray}
L_{1,2}^{x'_1}(t,t_0)&=&\sum_{x''_1}G_{1,2}^{1,x''_1}(t_0)G_{x''_1}^{x'_1}(t-t_0), \\
{\hskip -0.4cm }L_{1,2}^{x'_1,x'_2,x'_3}(t,t_0)&=&\sum_{x''_1,x''_2}G_{1,2}^{x''_1,x''_2}(t,t_0)G_{1,x''_1,x''_2}^{x'_1,x'_2,x'_3}(t,t_0).
\end{eqnarray}
Now, when we calculate the $\langle\tilde \sigma_n^z(t)\rangle$, each term in the pure state will contributes independently as the magnon states  are eigenstate for $\sigma_n^z$, but
only the terms with $L$ functions will contribute for the off-diagonal matrix element $\langle \tilde \sigma_n^+\rangle$. The expressions for these expectation values are very long, but
straightforward to evaluate. We have computed them numerically, and the results  for $F_n(t)$ and $Re O_n(t)$ are shown in Fig. 4(a) and Fig.4(b) respectively, with a representative values of the coupling strengths for $n=2,4,8$.
The excess of the local entropy for $n'$th qubit can be calculated from the eigenvalues for the reduced density matrix given by 
$\lambda_{\pm}=1\pm \sqrt{ \langle \tilde \sigma_n^z(t)\rangle^2
-\langle\tilde \sigma_n^+\rangle \langle \tilde \sigma_n^-\rangle }$. 
Similarly, for the state $\rho(t)$ with no QDP occurrence, we can calculate the local entropy. We have plotted the local
entropy detector function $D_n(t)$ in Fig. 4(c) for $n=2,4,8$. As can be seen from these figures, in this case all the detector functions shown have similar behaviour as in the case of even-only magnon
state cases with diagonal reduced density matrix. That is, the detector functions become nonzero after a waiting time, with the waiting time itself increasing with the distance of the measured
qubit from the QDP occurrence site. The waiting time $t_n^*$ is slightly different for the three different detector functions. But, we have seen that in this case also, the speed of the signal propagation is the same as calculated in the previous case, as the speed depends on the differential increase with $n$.

 \section{{XY model with transverse magnetic field}}
 
Another way of getting non trivial dynamics for the spin chain is to replace the bilinear Ising term with coupling strength $J_z$ in the last section by a linear transverse magnetic field term. 
Thus, we will have  a quantum spin chain of N spins interacting with their nearest neighbours in the XY plane, along with an external magnetic field in transverse direction. The Hamiltonian for this case is
given by,
  \begin{equation}
H = \sum_{i}(J_x{\sigma}^x_{i}{\sigma}^x_{i+1}+J_y{\sigma}^y_{i}{\sigma}^y_{i+1}) + h\sum_{i}{\sigma}^z_{i}.
\end{equation} 
This Hamiltonian can be exactly diagonalised and entire eigenvalue spectrum can be found for periodic boundary by employing Jordan-Wigner transformation\cite{ref6} of spin $1/2$ operators to spinless fermionic operators. The ground state  exhibits a quantum critical behaviour, for the isotropic case of $J_x=J_y$ for all values of the magnetic field strength, and for the anisotropic case for
$h=J_x+J_y$. In addition to the unitary dynamics of the above Hamiltonian, we will have a QDP ($\hat n=\hat z$ for simplicity) occurring at the first spin at an epoch time $t_0$ like
in the previous sections.

It is easy to see that the three different terms in the above Hamiltonian do not commute with each other. The dynamics is similar to the Ising dynamics we have discussed before, when
only one of the coupling constants is nonzero.  For example, let us take the limit $J_x=J_y=0$. Then the operator ${\sigma}^z_{i}(0)$ does not evolve with time for all sites. 
This implies $F(t) = 0$ for all sites independent of initial state. So, the detection of QDP is not possible. The case of $J_y=h=0$, or $J_x=h=0$, is similar to the discussion of Section II. However, for at least two of three parameters $J_x, J_y$ and $h$ nonzero the two terms in the Hamiltonian do not commute and non trivial dynamics can be observed. In general, it is expected that the speed should depend on the ratios $J_y/J_x$ and $h/J_x$.
 
 We map spin-1/2 operators in the Hamiltonian to Fermionic creation and annihilation operators by means of Jordan-Wigner transformation  \cite{ref6}. The mapping is given by,
 \begin{equation}
\sigma_l^{+} = c^{\dagger}_le^{i\pi \sum_{m=1}^{l-1} c_m^{\dagger}c_m }.
 \end{equation}
 The Hamiltonian will have a bilinear form in terms of Fermionic creation and annihilation operators, which can be be brought to a diagonal form by doing a Fourier transformation, followed
 by a Bogoliubov transformation \cite{ref5}.
Fourier transforming the operators into momentum space, we define 
\begin{equation}
c_q={1\over \sqrt{N}}\sum e^{-iql}c_l.
\end{equation}
Here, the set of allowed momentum values are given by is $q=2\pi m/N$, with $ m = -(N-1)/2..-1/2,1/2..(N-1)/2$ for odd  $N$; and $m = -N/2..0..N/2$ for even $N$.
In terms of these momentum-space operators the Hamiltonian has a bilinear form with non-diagonal operators $c^{\dagger}_qc^{\dagger}_{-q}$ and similar terms.

To diagonalise the Hamiltonian we employ Bogoliubov-Valatin transformation in which new Fermionic creation and annihilation operators  
are formed as a linear combination of old operators, given as
\begin{equation}
\eta_{1q}=u_q c_{q}-i v_q c^\dagger_{-q},{~ ~} \eta_{2q}=-iv_q c_q +u_q c^{\dagger}_{-q}.
\end{equation} The expansion coefficients and the eigenvalues are given by,
\begin{equation}
u_q = \sqrt{{1\over 2}+{(J_x+J_y)\cos q + h\over |\omega_q|}},~~v_q = \sqrt{1-u_q^2},
\end{equation} \begin{equation}
\omega_q = 2\sqrt{[(J_x+J_y) \cos q +h]^2+[(J_x+J_y)\sin q]^2}.
\end{equation}
In terms of these new fermion operators, the Hamiltonian is diagonal, we have
\begin{equation}
H = \sum_{0<q<\pi} |\omega_q| (\eta^{\dagger}_{1q} \eta_{1q} - \eta^{\dagger}_{2q} \eta_{2q}).
\end{equation}

Now, to calculate the QDP detector function for any initial state $|\Psi_0\rangle$, we need to find the time-evolved state for $t>t_0$, or equivalently find the time-evolved operators. The
detector function is given by,
$$ F_n(t)= \langle \Psi_0|\tilde \sigma_n^z(t) -  \sigma_n^z(t)|\Psi_0\rangle = Tr \sigma_n^z [\tilde \rho(t) - \rho(t)].$$
Here, in the first equation we need the difference of the time-evolved operator $\tilde \sigma_n^z$ with the QDP occurring at the first site, and the time-evolved operator $\sigma_n^z$ without the QDP occurring. In the second equation, we need the time-evolved states $\tilde \rho(t)$ and $\rho(t)$ with and without QDP occurring respectively. It is easier to calculate
the time-evolved operators first, through the first equation, and then find expectation values in various initial states. However, for some initial states, it is easier to calculate the time-evolved
state. Using the time-evolved operators, we can rewrite the detector function as,
\begin{eqnarray}
{\hskip -0.4cm}F_n(t)&=& 2\langle\Psi_0|  P_0^\dagger (t_0) \sigma_n^z (t) P_0(t_0)|\Psi_0\rangle
\langle \Psi_0|P_0^\dagger (t_0) \sigma_n^z (t) +\sigma_n^z (t) P_0(t_0)|\Psi_0\rangle.
\end{eqnarray}
Here $P_0(t_0)=U_{0,t_0}^\dagger P_0 U_{0,t_0}$ is the time-evolved Kraus operator for the QDP, that involves the time evolution of $\sigma_1^z$. Now, both the spin operators of
the first qubit and $s$'th qubit do not commute with the Hamiltonian, and they involve the bilinear fermion operators $c_1^\dagger c_1$ and $c_n^\dagger c_n$ respectively. Now, the
time evolution of the fermion annihilation operator in momentum space can be calculated, we have
\begin{equation}
c_q(t) = (e^{-i\omega_qt}u_q^2+ e^{i\omega_qt} v_q^2) c_q - {q\over |q|} 2 u_q v_q \sin{\omega_q t} ~c_{-q}^\dagger.
\end{equation}
We can see above how the time evolution mixes the different operators. The detector function involves a product of three Pauli spin operators, that is a product of six fermion operators in
the real space. That is, it involves six sums over momenta values and eight different expectation values of  products of six fermion operators in momentum space, calculated in the initial
state. The expectation values of products of fermion operators can be straightforwardly evaluated using analog of Wick's theorem. However, there will be three sums over the momentum variables that can be carried out numerically.
\begin{figure*}
 \begin{tabular}{cc}
 \includegraphics[width=0.5\textwidth]{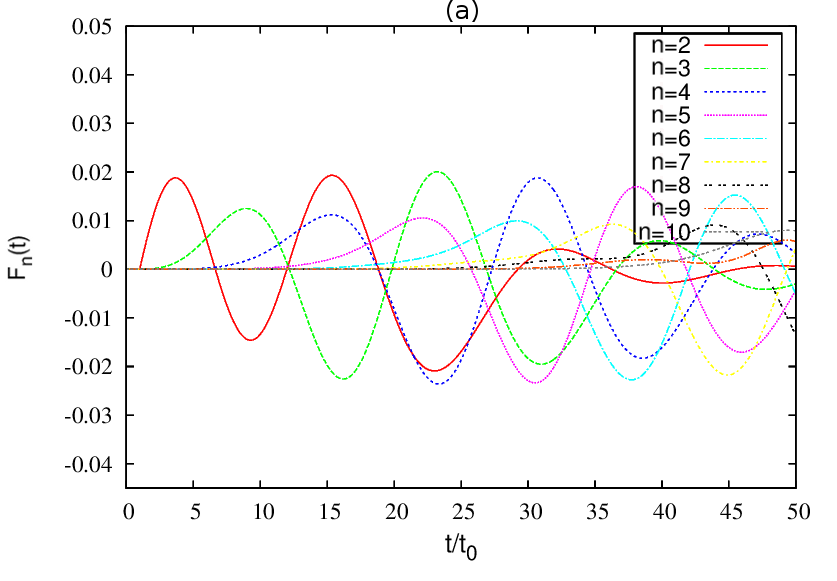}&
 \includegraphics[width=0.5\textwidth]{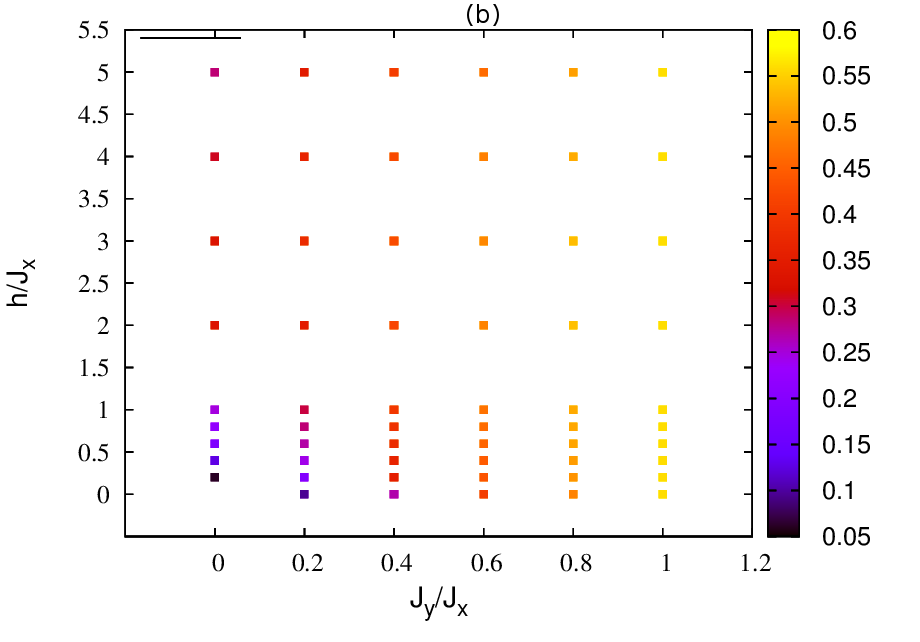}\\ 
 \includegraphics[width=0.5\textwidth]{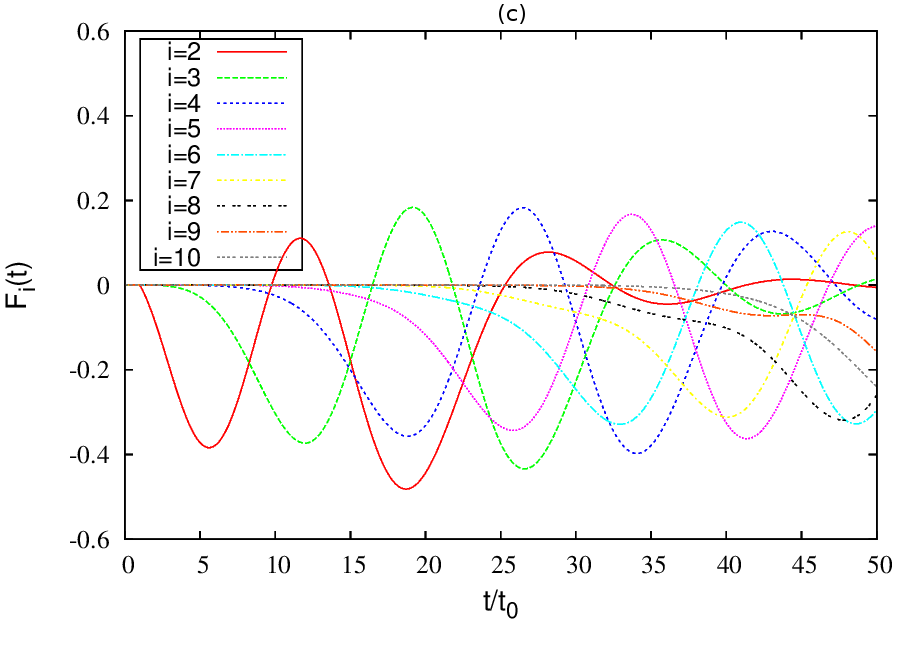}& \includegraphics[width=0.5\textwidth]{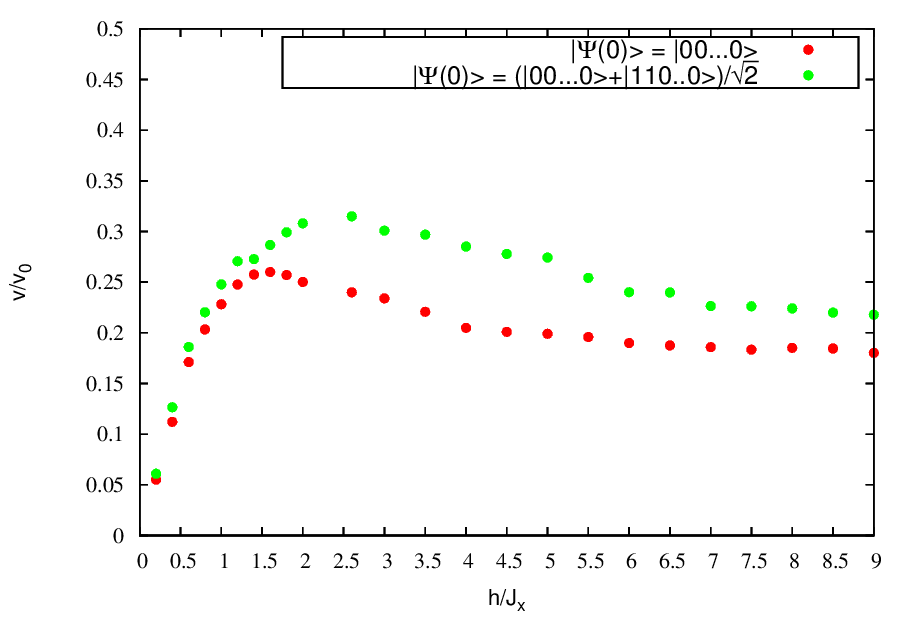}
  \end{tabular}
 \caption{\label{fig:fig_3}{\small For the XY dynamics with a transverse magnetic field with open boundary conditions, and the QDP with $\hat n=\hat z$: ( a) $F_n(t)$ is plotted with $t/t_0$ for the initial state $|000...0\rangle$ with interaction strengths $J_x=0.7$,$J_y=0.3$ and $h=1.0.$
  (b)  The density plot of the speed of propagation for various values of $J_y/J_x$ and $h/J_x$, for the initial state $|000...0\rangle+|110..0\rangle/\sqrt{2}$. The speed is maximum for the quantum critical  case of $J_x = J_y$ independent of  $h$.
(c) The Detector function $F_n(t)$ is plotted with time $(t/t_0)$ for the initial state $(|000...0\rangle+|110..0\rangle)/\sqrt{2}$, with interaction strengths $J_x=0.7$,$J_y=0.3$ and $h=1.0$ 
(d) The speed has been plotted with $h/J_x$ for for the two initial states  $|000...0\rangle+|110..0\rangle/\sqrt{2}$ and $|000...0\rangle$. Here, the reference
speed is $v_0=10 |J_x|a/\hbar$ in terms of $a$, the nearest-neighbour separation between the qubits.
 
 }}
\end{figure*}

For some simple initial states, the detector function can be calculated from the time-evolved state, as we outline below.
Let us consider a simple initial state $|00...00\rangle$ with no entanglement. This state can evolve into a complicated state with multipartite entanglement structure\cite{arul} through time evolution. We
can rewrite the initial state as,
\begin{equation}
|\Psi(0)\rangle = \prod_{q>0}|0\rangle_q| 0\rangle_{-q} = \prod_{q>0} (v_q \eta^{\dagger}_1- u_q \eta^{\dagger}_2)|vac\rangle,
\end{equation}
where the vacuum state for the $\eta$ operators is  $|vac\rangle =i|0\rangle_q|1\rangle_{-q}$, as a linear combination of the eigenstates of the Hamiltonian. 
 n the new basis Hamiltonian being diagonal each $q$ mode has independent time evolution. The state after a time $t$ becomes,
\begin{equation}
|\Psi(t)\rangle= \prod_{q>0}(v_q e^{i\omega_qt} \eta^{\dagger}_1- u_q e^{-i\omega_qt}\eta^{\dagger}_2)|vac\rangle\equiv \prod_{q>0} |\phi(t)\rangle_q,
\end{equation}
where we have defined  $|\phi(t)\rangle_q$, a time-evolved linear combination of eigenstates for each momentum $q$.
Just after the QDP occurs at the first site the state is given by Eq(44). Here the Kraus operators $P_0$ and $P_1$ are written in terms of the new set of fermionic creation and annihilation operators formed via Bogoliubov transformation; $P_0 = \frac{1}{N} \sum_{q_1,q_2} c^{\dagger}_{q_1} c_{q_2} e^{i(q_1-q_2)} = \frac{1}{n} \sum_{q_1,q_2} (u_{q_1} \eta_{1}^{\dagger}+v_{q_1}\eta_{2}^{\dagger})(u_{q_2} \eta_{1}+v_{q_2}\eta_{2}) e^{i(q_1-q_2)} $ for $q>0$;  where as for $q<0$ it is give by $ P_0 = \frac{1}{N} \sum_{q_1,q_2} (v_{q_1} \eta_{1}^{\dagger}-u_{q_1}\eta_{2}^{\dagger})(v_{q_2} \eta_{1}-u_{q_2}\eta_{2}) e^{i(q_1-q_2)} $,  and $P_1 = (\mathbf{1} - P_0)$. The state $\tilde \rho(t_0^+)$, just after the QDP,  will involve two pure states, and is given by,
\begin{equation}
 \tilde{\rho}(t_{0^{+}}) = |\tilde\Psi_1(t_{0^{}})\rangle\langle \tilde\Psi_1(t_{0^{}})|+|\tilde\Psi_2(t_{0^{}})\rangle\langle \tilde\Psi_2(t_{0^{}})|.
 \end{equation}
 Here,
 \begin{eqnarray}
{\hskip -1cm} |\tilde\Psi_1(t_{0^{}})\rangle = \frac{1}{N}[\sum_{q_1,q_2} |\phi_+(t_{0^{}}) \rangle_{q_1} |\phi_-(t_{0^{}}) \rangle_{q_2} \prod_{q \neq q_1,q_2}|\phi (t_{0^{}})\rangle_q  +
 \sum_{q_1} |\phi_1(t_{0^{}}) \rangle_{q_1} \prod_{q \neq q_1}|\phi (t_{0^{}})\rangle_q],\nonumber \\
   |\tilde\Psi_2(t_{0^{}})\rangle =|\Psi(t_{0^{}})\rangle -|\Psi_1(t_{0^{}})\rangle,
 \end{eqnarray} 
 where we have used three time-evolved states for a momentum value,
   \begin{equation}
   {\hskip -1.3cm} |\phi_+(t_{0^{}}) \rangle_{q} = c^{\dagger}_q e^{iq} |\phi(t_0)\rangle_q ,~    |\phi_-(t_{0^{}}) \rangle_{q} = c_q e^{-iq} |\phi(t_0)\rangle_q , ~
 |\phi_1(t_{0^{}}) \rangle_q = c^{\dagger}_qc_q |\phi(t_0)\rangle_q.
 \end{equation}
 Evolving further the state for a time $t>t_0$ we get, 
  \begin{equation}
 \tilde{\rho}(t) = |\tilde\Psi_1(t)\rangle\langle \tilde\Psi_1(t)|+|\tilde\Psi_2(t)\rangle\langle \tilde\Psi_2(t)|,
 \end{equation}
where the two time-evolved pure states are given as,
  \begin{eqnarray}
{\hskip -0.5cm} |\tilde\Psi_1(t)\rangle = \frac{1}{n}[\sum_{q_1,q_2} |{\phi}_+(t) \rangle_{q_1} |{\phi}_-(t) \rangle_{q_2} \prod_{q \neq q_1,q_2}|\tilde{\phi }(t)\rangle_q \nonumber +
 \sum_{q_1} |\tilde{\phi}_1(t) \rangle_{q_1} \prod_{q \neq q_1}|\tilde{\phi} (t)\rangle_q],\nonumber \\
   |\tilde\Psi_2(t)\rangle = |\Psi(t)\rangle -|\tilde\Psi_1(t)\rangle.
 \end{eqnarray} 
 In the above we have introduced two more time-evolved states for a momentum value, given by
   \begin{eqnarray}
 {\hskip -1.5 cm} |\tilde{\phi}(t) \rangle_{q} = (u_q e^{i\omega_qt} \eta^{\dagger}_2+v_q e^{-i\omega_qt}\eta^{\dagger}_1)|vac\rangle, 
      |\tilde{\phi}_1(t) \rangle_{q} = (u_q e^{i\omega_qt} \eta^{\dagger}_1+v_q e^{-i\omega_qt}\eta^{\dagger}_2)|vac\rangle.
  \end{eqnarray}
    The time evolved state at a time $t$ are given in Eq.49 and Eq. 45, with and without the QDP occurring. We can calculate the detector function, using these states, we have 
 
 \begin{equation}
{\hskip -2.3 cm} F_n(t) ={ \frac{4}{N}}  {\rm Re} {\Bigg \{} \sum_{q_1,q_2} e^{i(q_1-q_2)n} [ \langle \Psi_1(t)| c_{q_1}^{\dagger} c_{q_2}|\Psi_1(t)\rangle \\- \langle \Psi_1(t)| c_{q_1}^{\dagger} c_{q_2}|\Psi(t)\rangle ] {\Bigg \}  }+ 2\langle\Psi_1(t)|\Psi_2(t)\rangle.
 \end{equation}
 Now, the expectation value $ \langle \Psi_1(t)| c_{q_1}^{\dagger} c_{q_2}|\Psi_1(t)\rangle$ is straightforward to evaluate, and similarly the other matrix elements. However, the detector function has a long expression with two sums over the momentum variables, that can be calculated numerically.
 In this case, we have calculated the quantity $F_n(t)$ by calculating the sums over $q_1$ and $q_2$ numerically, and it has been seen that the result matches with the exact diagonalisation
results. Fig. 5(a) shows $F(t)$ for the initial state $|00..0\rangle$. However, for a general entangled state the calculation of the detector function is quite difficult. 
 
We have seen that in the case of XY model with transverse magnetic field at least two of
the three parameters $J_x$, $J_y$ and $h$
should be non zero to obtain non trivial dynamics. It has been seen that the speed
of the signal depends on these three parameters. When the magnitude of the magnetic
field is zero we expect maximum speed when $J_x = J_y$; which is also a quantum anisotropy phase transition point between two ferromagnetic phases. Numerical calculations
confirm this. Also when magnetic field is increased the speed also increases but can
not be increased beyond a certain value of the field. The Fig.5(b) shows the dependence of the speed on these three parameters, where the reference speed is $v_0=10 |J_x|a/\hbar$,
analogous to the earlier discussion, the speed of the signal is of the order $10^4 M/sec$. As compared to the previous case of Heisenberg
dynamics that conserves the z-component of the total spin, in the present case of non conserving dynamics, we see a stronger dependence of the detection speed on the parameters
$h/J_x$ and $J_y/J_x$. This is expected as we have more parameters and non conservation. In this case we have studied only a simple initial state analytically, as initial states with
entangled pair of spins are quite difficult to handle for this model Hamiltonian. We take a numerical approach for studying other initial states discussed below.

Figure 5(c) shows $F(t)$ as a function of time for the initial state $(|00..0\rangle+|110..0\rangle)/\sqrt{2}$. The actual dynamics is different from that of the direct product state considered in Figure 5(b) and Figure 5(d) show the variation of speed with the magnetic field strength for the two initial states $(|00..0\rangle+|110..0\rangle)/\sqrt{2}$ and $|00..0\rangle$ when $J_y = 0$.   
Though the two states differ in terms of the initial entanglement (between the first two spins), since the dynamics will generate a variety of entanglement distributions,  the speed of
the QDP signal is expected to be the same. The signal speed reach an optimal value as a function of the magnetic field strength, and after that it decreases slowly to zero for large
fields. In the limit of a large magnetic field, the interaction terms in the Hamiltonian (Eq.38) are insignificant and can be ignored. Since, the qubits evolve independently in this case, the signal speed goes to zero. making it impossible to detect the signal. 

In all the cases studied till now, we have used integrable systems for the Hamiltonian dynamics for the unitary evolution of the state. In the case of non-integrable systems, the behaviour
of the wait times could be quite different. To this end, we consider the case of applying a longitudinal field also in addition to the transverse field for the XY dynamics that we have studied above. 
The presence of both the transverse field $h$ (along z direction as shown in Eq. 38) and a longitudinal field $h'$ (along x direction that couple to $\Sigma \sigma_i^x$), will
render the system to be non-integrable, which has been studied in the context of quantum chaotic behaviour \cite{Prosen}. The system is no longer exactly solvable using the Jordan-Wigner 
transformation unlike the case of $h' =0$ we studied above, thus eliminating a analytical tool to analyse the dynamics and getting the equivalent of Eq.53 and Eq.55.
We can, however, proceed numerically for getting the energy spectrum and the eigenfunctions that are needed to compute the time evolved state, and the detector function. We have
plotted the results for the detector function in Fig.6, for the initial state $|00..0\rangle$, for a set of representative values of the coupling strengths (the same set used for Fig.5a with
an additional longitudinal field of the same magnitude) for an open finite chain with ten qubits. It can be seen that the case of $h' \ne 0$ (non-integrable system) the behaviour is very different from the case of $h'=0$ (integrable), the function becomes nonzero for all qubits at the same time. Here, there is no difference in the waiting time as we increasing the distance from the QDP site, implying that we
can define a signal speed. We have seen a similar situation in for the long-ranged Ising dynamics in Fig.2c, where it is due to the direct interaction of the further qubits with the first qubit.
But here, there seems to be no waiting time for far away qubits, though we have only nearest-neighbour interactions.
This implies that the fact that we have a non-integrable dynamics is crucial for this behaviour, and it is difficult to explain this conclusively with a simple example, with no analytical insight. 
It will be interesting to investigate in detail various other initial states, and other non-integrable models, to get an insight into this waiting time behaviour of the QDP signal.

\begin{figure}
\center{
 \includegraphics[width=0.55\textwidth]{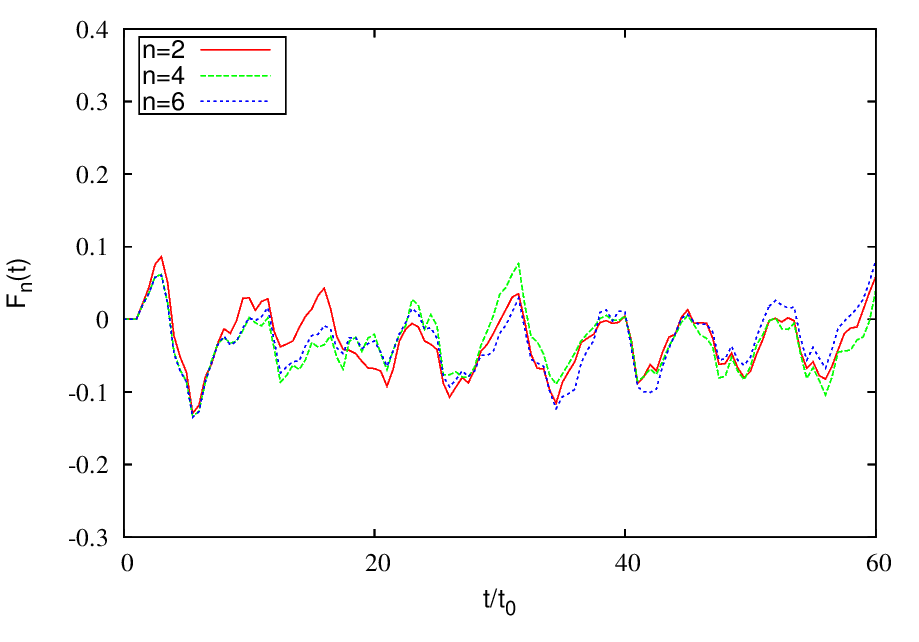} }
\caption{Non-integrable Hamiltonian dynamics: XY model with transverse and longitudinal fields with open boundary conditions. $F_n(t)$ is plotted with $t/t_0$ for the initial state $|000...0\rangle$ with interaction strengths $J_x=0.7$,$J_y=0.3$ and the transverse field $h=1.0.$ and a longitudinal field $h'=1.0$ for a finite chain with ten qubits. It can be seen that the detector function
for different qubits does not show different waiting times, unlike the integrable case shown in Fig.5 (with $h'=0$).
}
\end{figure}
\section{{Conclusions}} 

We investigated a model of decoherence in an interacting spin chain resulting from a Quantum dynamical process on a single site. We have discussed the possibility of detection of the QDP signal from a different site, for different initial states and different Hamiltonian dynamics of the system. In this paper we have presented analytical calculations along with
numerical results for specific initial states, viz. polarized direct product state, Hamiltonian eigenstates, pairwise entangled state, globally entangled GHZ state etc.

For the Ising dynamics, we have shown that the effect of QDP can not propagate through the chain beyond the second site for nearest neighbour interactions.  For the case of long-ranged interactions, the effect reaches all the sites instantaneously; which can be intuitively understood.
Thus, it is not possible to define a speed of signal propagation for this case. In the case of entangled initial states, the QDP signal cannot be detected from a site which is entangled with
the first site initially.

We have shown that it is possible for the QDP signal to reach further sites, within near-neighbour interacting models for the Heisenberg and XY models.
In the case of Heisenberg model the speed is directly proportional to the interaction strength between the spins, which is as expected. Also, for the anisotropic case the speed is maximum when the anisotropy parameter $J_z = 0$. For anisotropic XY model with the magnetic field in transverse direction, it is seen that the speed depends on both the ratios $J_y/J_x$ and $h/J_x$. Maximum speed is observed for the quantum critical case of $J_x=J_y$ for any value of $h$. However, the speed is not maximum for the quantum critical case of $J_y=0$ and $h=J_x$.
In these case, any entanglement present in the initial state does not change the QDP signal propagation and the detection.

The dynamics in general depends on the initial state and the nature of interaction between the spins. The speed of the signal depends only the parameters of the Hamiltonian not on the initial state. The speed is similar for an unentangled or entangled state. However, if the initial state is an eigenstate of the Hamiltonian and the measurement operator, it is not possible to detect the signal. We have also seen that for non-integrable systems we cannot define a signal speed.
Our method is based on a simplistic model of QDP, where the environment interacts (ex. Quantum measurement) with the system instantaneously through only one spin and that gives rise to decoherence in a many body spin system. However, there can be other complicated models for the same. We have only discussed for one dimensional spin chains, the dynamics for higher dimensional cases can be more complicated.

        \end{document}